\documentclass[5p,sort&compress]{elsarticle}
\pdfoutput=1
\usepackage{amsmath}
\usepackage{amssymb}
\usepackage[mathlines]{lineno}
\usepackage{subfigure}
\usepackage{cuted}
\usepackage{booktabs}
\usepackage{threeparttable}
\usepackage{upgreek}
\usepackage{xcolor}

\newcommand{\neth}{n_{\rm e,th}}
\newcommand{\phn}{\hphantom{0}}
\newcommand{\chead}[1]{\multicolumn{1}{c}{\ensuremath{#1}}}

\graphicspath{{figures/}} 

\journal{Astroparticle Physics}

\newcommand{\eqnref}[1]{Eq.~\ref{#1}}
\newcommand{\eqnrefii}[2]{Eqs~\ref{#1} and~\ref{#2}}
\newcommand{\eqnrefiii}[3]{Eqs~\ref{#1}, \ref{#2} and~\ref{#3}}
\newcommand{\secref}[1]{Sec.~\ref{#1}}
\newcommand{\secrefii}[2]{Secs~\ref{#1} and~\ref{#2}}
\newcommand{\appref}[1]{App.~\ref{#1}}
\newcommand{\figref}[1]{Fig.~\ref{#1}}
\newcommand{\figrefii}[2]{Figs~\ref{#1} and~\ref{#2}}
\newcommand{\tabref}[1]{Table~\ref{#1}}

\begin{document}

\begin{frontmatter}

\title{An updated estimate of the cosmic radio background and implications for ultra-high-energy photon propagation}

\author[manchester]{I.~C.~Ni\c{t}u}
\author[manchester]{H.~T.~J.~Bevins}
\author[manchester]{J.~D.~Bray}
\author[manchester]{A.~M.~M.~Scaife}

\address[manchester]{JBCA, Dept.\ of Physics \& Astronomy, Univ.\ of Manchester, Manchester M13 9PL, UK}

\begin{abstract}
We present an updated estimate of the cosmic radio background (CRB) and the corresponding attenuation lengths for ultra-high energy photons.  This new estimate provides associated uncertainties as a function of frequency derived from observational constraints on key physical parameters.  We also present the expected variation in the spectrum of the CRB as a function of these parameters, as well as accounting for the expected variation in spectral index among the population of radio galaxies.  The new estimate presented in this work shows better agreement with observational constraints from radio source-count measurements than previous calculations.  In the energy regime where we expect cosmogenic photons dominantly attenuated by the CRB, our calculation of the attenuation length differs from previous estimates by a factor of up to 3, depending on energy and the specific model for comparison.  These results imply a decrease in the expected number of cosmogenic photons with energies $\sim 10^{19}$--$10^{20}$\,eV.
\end{abstract}


\end{frontmatter}

\section{Introduction} \label{sec:intro}

Ultra-high-energy (UHE; $>$10$^{18}$\,eV) photons have the potential to constitute a powerful probe of the most energetic processes in the Universe, if they can be detected and distinguished from the dominant background of ultra-high-energy cosmic rays (UHECRs).  Although the scope for UHE photons to originate directly from exotic physical processes (e.g.\ \citep{ellis2006,aloisio2015}) has been diminished by experimental limits~\citep{aab2017b,abbasi2019}, the interaction of UHECRs with the cosmic microwave background (CMB) through the Greisen-Zatsepin-Kuzmin (GZK) effect~\citep{greisen1966,zatsepin1966} is confidently expected to produce a baseline photon flux (e.g.\ \citep{taylor2009,hooper2011}).  The spectrum of these UHE photons can act as a probe of the composition of UHECRs~\citep{hooper2011} and, unlike UHECRs, they propagate without deflection by magnetic fields, so they can potentially also provide directional information about the objects that accelerate particles up to these energies.

UHE photons are, however, attenuated by interactions with background photon fields, where the dominant process is interaction of two photons to produce an electron-positron pair.  This process is a long-established prediction of quantum electrodynamics~\citep{Breit1934}.   While pair-production from the interaction of three or more photons has been observed in the laboratory ~\citep{Burke1997}, approaches for direct detection of the two-photon process are under review~\citep{Ribeyre2016}. The cross-section of this interaction is maximised when the combined energy of the two photons, in their centre-of-mass frame, is close to the threshold energy defined by the rest mass-energy $m_{\rm e} c^2$ of the electron and positron.  Consequently, a field of background photons with a characteristic energy~$\varepsilon$ will most strongly attenuate incident photons with a corresponding energy $E \sim m_{\rm e}^2 c^4 / \varepsilon$.  Astrophysically, the dominant background photon field throughout most of the Universe is the CMB, which has a characteristic thermal spectrum with a mean energy of $6.34 \times 10^{-4}$\,eV~\citep{SpectrumUniverse}.  For photons with energies around $10^{15}$\,eV, where the resulting attenuation is strongest, this results in an attenuation length of $<10$\,kpc~\citep{Protheroe1996}, completely preventing the detection of extragalactic sources and significantly attenuating even sources within our own Galaxy.

In contrast, more energetic photons are only weakly attenuated by the CMB, as the centre-of-mass energy of a UHE photon and a CMB~photon is far greater than the threshold for the pair-production process.  The resulting attenuation lengths are on scales of Mpc, allowing even extragalactic sources to be detected.  In this regime the attenuation of UHE photons will instead be dominated by extragalactic interactions with background photons less energetic than the CMB.  Known as the extragalactic radio background (EGRB), universal radio background (URB) or cosmic radio background (CRB), this lower-energy photon field is expected to dominate attenuation of photons with energies above a threshold $\sim 10^{19}$\,eV~\citep{Protheroe1996}.  Consequently, a precise understanding of the flux and spectrum of the CRB will determine the detectability of extragalactic sources of UHE photons.

Measuring the CRB is difficult primarily because of the ionosphere, which distorts and attenuates radio observations at low radio frequencies, and additionally due to foreground emission from the Galaxy, which must be subtracted in order to isolate the extragalactic component.  Estimates of the CRB have been reported by \citet{clark1970}, who used satellite radio measurements to bypass the ionosphere, and subtracted a model of the Galactic foreground; and \citet{Protheroe1996}, who modelled the radio emission from populations of star-forming galaxies (SFGs) and radio galaxies (RGs) to determine the resulting background flux.  Monte Carlo simulations of UHE particle propagation~\citep{armengaud2007,batista2016,aloisio2017} incorporate one or both of these to allow estimation of the UHE photon flux (e.g.\ \citep{heiter2018,batista2019}).

In this work, we present an updated estimate of the CRB.  We refine the approach of \citet{Protheroe1996} by 
(i)~considering both synchrotron and free-free emission for SFGs in order to constrain the high-frequency behaviour of the CRB more precisely,
(ii)~using a more realistic resolved-disk model for SFGs based on M51,
(iii)~using a more representative model for RGs based on the 3CRR survey, including a distribution of spectral indices,
(iv)~propagating errors from key physical parameters to provide an estimate of the uncertainty in our prediction of the CRB as a function of frequency,
(v)~using a full analytic treatment of synchrotron self-absorption to constrain the low-frequency behaviour of the CRB more precisely, and
(vi)~using a model for galaxy evolution based directly on radio source counts in an updated $\Lambda$CDM cosmology.

The structure of the paper is as follows: in \secrefii{sec:SFGs}{sec:RGs} we describe the procedures by which we derive our estimates for the CRB contributions from SFGs and RGs respectively.  In \secref{sec:comparison} we explicitly compare our results to those of \citet{Protheroe1996} as well as recent observational constraints from radio source-count data, while \secref{sec:disc} contains a discussion of the uncertainties in our calculation.  In \secref{sec:intlen} we discuss the implications of our results for the attenuation and observation of UHE photons, and in \secref{sec:conc} we summarise our conclusions.

Throughout this work we assume a $\Lambda$CDM cosmology with $H_0 = 70\,$km/s/Mpc, $\Omega_m = 0.31 \pm 0.01$, $\Omega_\kappa = 0.005$ and $\Omega_\Lambda = 0.685 \pm 0.01$~\citep{Planck2015}.

\section{Star-forming galaxies}
\label{sec:SFGs}

Following \citet{Protheroe1996}, we derive an estimate of the CRB using a combination of the integrated radio emission from the cosmological populations of star-forming or ``normal'' galaxies (SFGs; this Sec.) and radio galaxies (RGs; \secref{sec:RGs}).  SFGs are galaxies undergoing a period of active star formation, in which gas is ionised by massive young stars and supernovae inject energetic electrons into the interstellar medium (ISM).  SFGs are not individually high-luminosity radio sources, but they are numerous, and their population is expected to dominate source counts of low-flux-density objects in radio surveys~\citep{mancuso2015}.

The dominant mechanisms by which SFGs contribute to the CRB are free-free emission from thermal electrons in the ionised ISM and synchrotron radiation from high-energy electrons in the galactic magnetic field.  To quantify these, we take the well-studied nearby galaxy M51 as a representative SFG, using observational data to constrain the relevant physical parameters and estimate its total radio luminosity~$L_\nu$ as a function of frequency~$\nu$ (\secref{sec:SFGtyp}).  By integrating a population of similar SFGs over cosmological history, we can then calculate their total contribution to the CRB (\secref{sec:SFGtot}).

\subsection{The typical luminosity spectrum of SFGs}
\label{sec:SFGtyp}

To calculate the free-free emission spectrum (see \appref{app:radiation}), we require the number density and temperature of thermal electrons in our exemplar SFG.  Previous measurements of the thermal-electron number density for M51 give the radial profile~\citep{Berkhuijsen1997}
 \begin{equation}
  \neth = 
   \begin{cases}
    0.11\pm 0.03\,\textnormal{cm}^{-3}              & r < 4.8\,\textnormal{kpc} \\
    0.06\pm 0.01\,\textnormal{cm}^{-3}              & 4.8 < r < 7.2\,\textnormal{kpc} \\
    0.013\pm 0.01\,\textnormal{cm}^{-3}             & 7.2 < r < 9.6\,\textnormal{kpc} \\
    0.004~^{+0.008}_{-0.003}\,\textnormal{cm}^{-3}  & r > 9.6\,\textnormal{kpc} \\
   \end{cases}	
  \label{BerkElecDen}
\end{equation}
where $r$ is the radial distance from the centre of the galaxy.  The uncertainties in $\neth$ are propagated through the free-free emission and absorption coefficients as per \eqnrefii{kap_f}{eps_f}.  We note that for electron densities in this range we are securely above the low-frequency cut-off described in \eqnref{nuplasma}.  We take the electron temperature in M51 to be $T_{\rm e} \approx 10\,000$\,K~\citep{Downes1980}; the effect of varying this assumption is discussed in \secref{sec:disc:vare}.

The synchrotron emission spectrum is a function of the energy density of high-energy electrons and the magnetic field strength.  Following~\citep{Protheroe1996}, we adopt a three-branch model for the electron energy density in SFGs~\citep{Nath1994,Protheroe1996},
 \begin{equation}
  n_{\rm e}(E) = n_0
   \begin{cases}
    (E/400\,\textnormal{MeV})^{-p}                       & E > 400\,\textnormal{MeV} \\
    (E/400\,\textnormal{MeV})^{-1.8}                     & 50 < E < 400\,\textnormal{MeV} \\
    (50\,\textnormal{MeV}/400\,\textnormal{MeV})^{-1.8}  & E < 50\,\textnormal{MeV}, \\
   \end{cases}
  \label{ne}
 \end{equation}
where $p$ is the energy spectral index and $n_0$ is a proportionality constant. Note that any radial dependence is contained in $p$, while $n_0$ is taken to be radially independent.  The shape of $n_{\rm e}(E)$ below~$400~\textnormal{MeV}$ is modified by ionisation and Bremsstrahlung losses~\citep{Strong1996, Pacho}.  The cut-off at 50\,MeV corresponds to the cyclotron radius of electrons being small enough that the electrons gain and lose energy continually by interaction with the plasma~\citep{Bell1978,Lesch1990}.   In this instance the electrons have a cyclotron frequency equal to the plasma frequency.  While they continue to emit radiation their cyclotron frequency remains constant and driven by the bulk motion of the plasma.

 It is worth mentioning that recent observations by Voyager 2 of the electron energy density of our galaxy~\citep{stone2019} are mostly consistent with the power-law behaviour described in Eq.~\ref{ne}.  See~\secref{sec:disc_SFG} for a more detailed discussion of the Voyager 2 results and their application in this analysis.

We estimate values for the magnetic field strength~$B$, the energy spectral index~$p$ and the proportionality constant~$n_0$ using observational data for M51.  This galaxy is located at a distance of \mbox{$d=7.6$\,Mpc~\citep{Ciardullo2002}} and has an inclination \mbox{$i_{\rm M51}=-20\,$deg~\citep{Tully1974}}.  

We adopt the usual power-law approximation for the frequency-dependent radio flux density:   \begin{equation}
  S_{\nu}^{\rm obs} = C \, \nu^\alpha
  \label{S_alpha}
 \end{equation}
where $\alpha$ is the spectral index and $C$ is a frequency-inde\-pendent constant of proportionality.  The spectral index of the electron energy distribution~$p$ in \eqnref{ne} is then related to the spectral index of the intensity, $\alpha$, by
 \begin{equation}
  p = -2 \alpha + 1.
  \label{p}
 \end{equation}

\begin{figure}
	\centering
	\includegraphics[width=\linewidth]{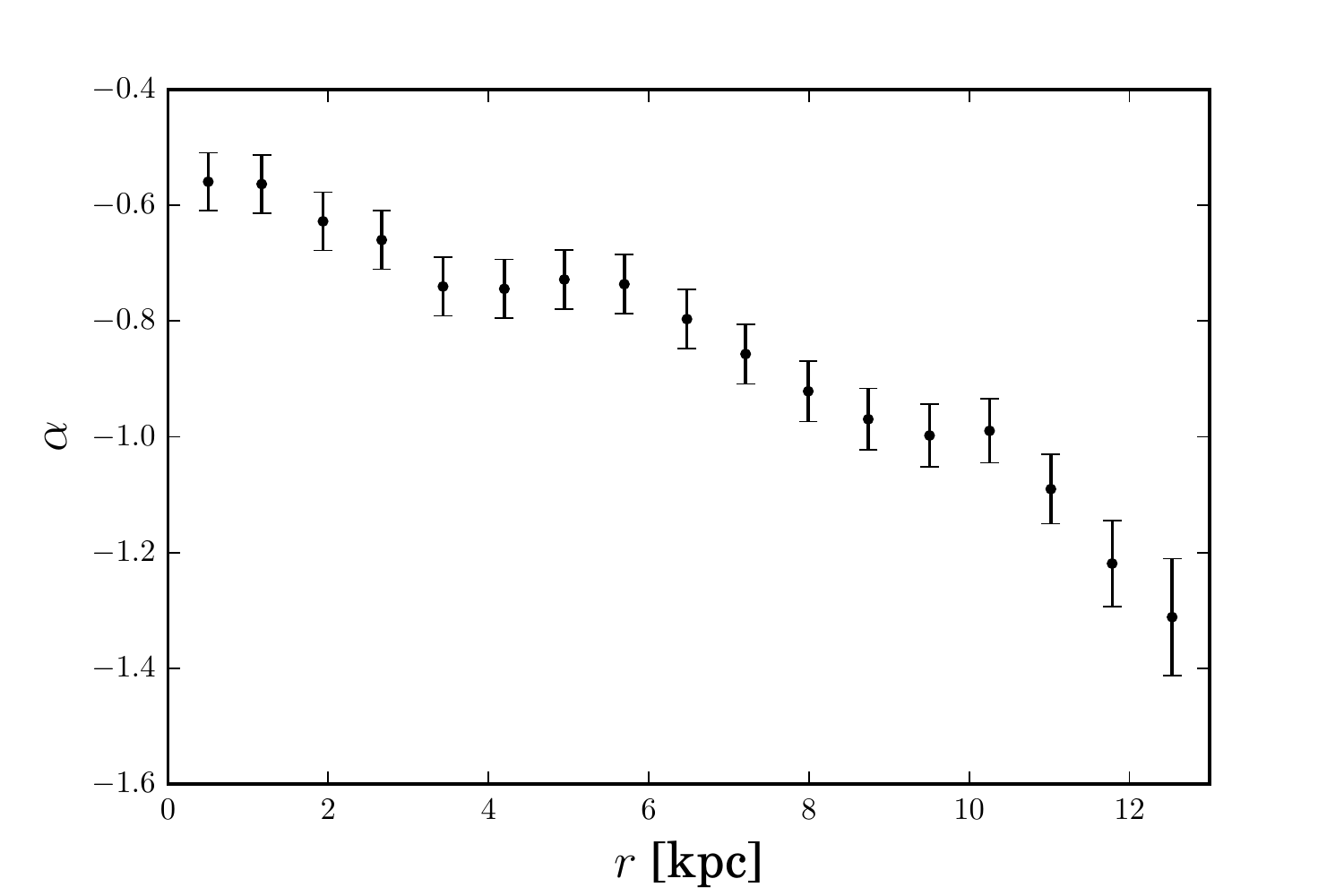}
	\caption{Spectral index of radio emission from the galaxy M51, as a function of radius.  This was obtained from the LOFAR and VLA observations of \citet{Mulcahy2014}.}
	\label{fig:alphavsR}
\end{figure}

\begin{figure}
	\centering
	\includegraphics[width=\linewidth]{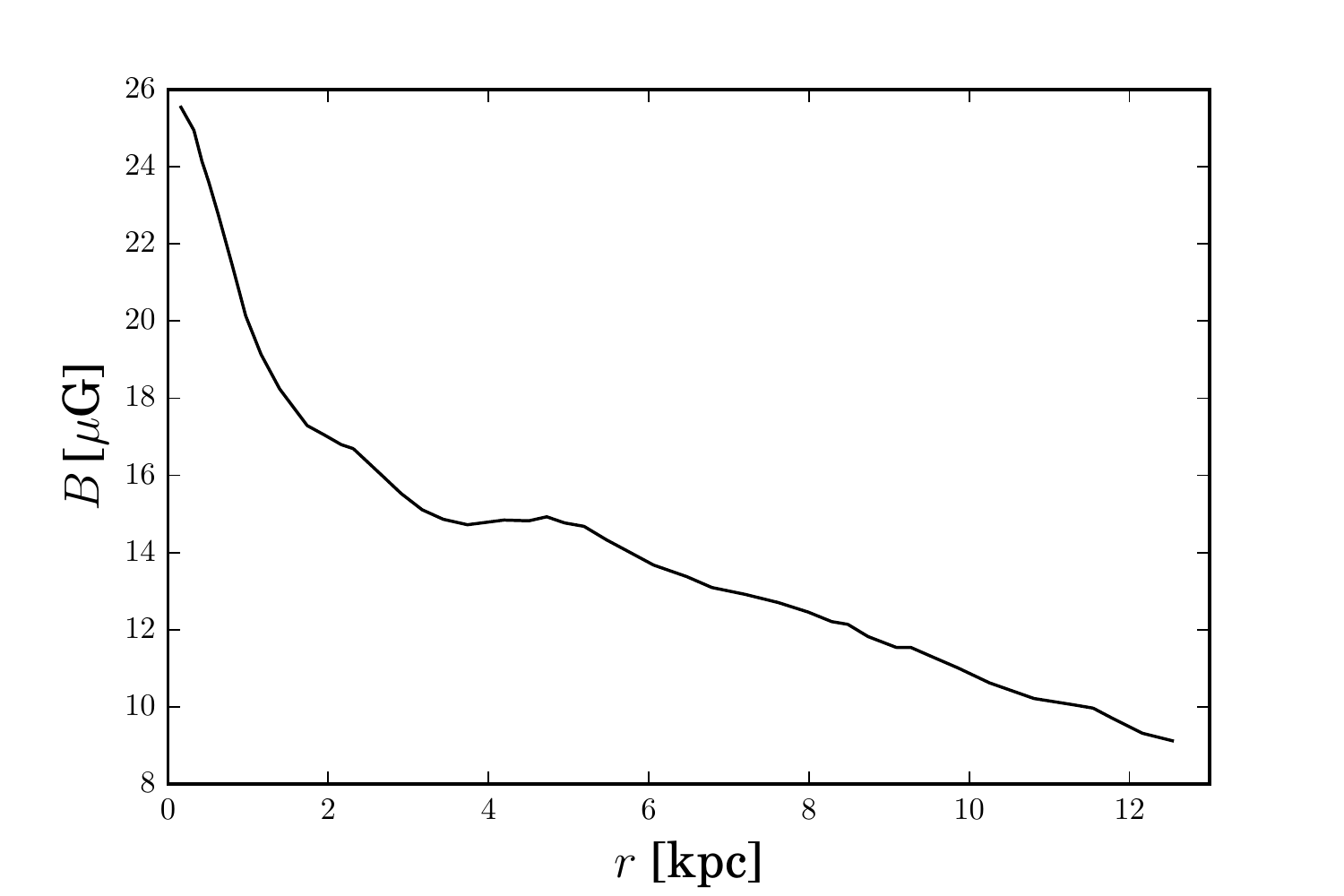}
	\caption{Magnetic field in the galaxy M51 as a function of radius, after Fig.~10 of \citet{Mulcahy2014}.  The associated uncertainties are $\pm 5$\%.}
	\label{fig:BvsR}
\end{figure}

We compute values of $C(r)$ and $\alpha(r)$ as a function of galactic radius $r$ with associated uncertainties (see \figref{fig:alphavsR}) in the interval from 151\,MHz to 1.4\,GHz using radially-resolved data from~\citep{Mulcahy2014}.  The uncertainties on these flux density measurements include contributions from both thermal noise and instrumental calibration effects.  The spacings between consecutive radial measurements are approximately equivalent to the telescope beam size, such that it is reasonable to assume that these uncertainties are uncorrelated.  The magnetic field strength of M51 at the same radial positions was adapted from the measurements of \citet{Mulcahy2014} and is shown in \figref{fig:BvsR}.

\begin{figure}
	\centering
	\includegraphics[width=\linewidth]{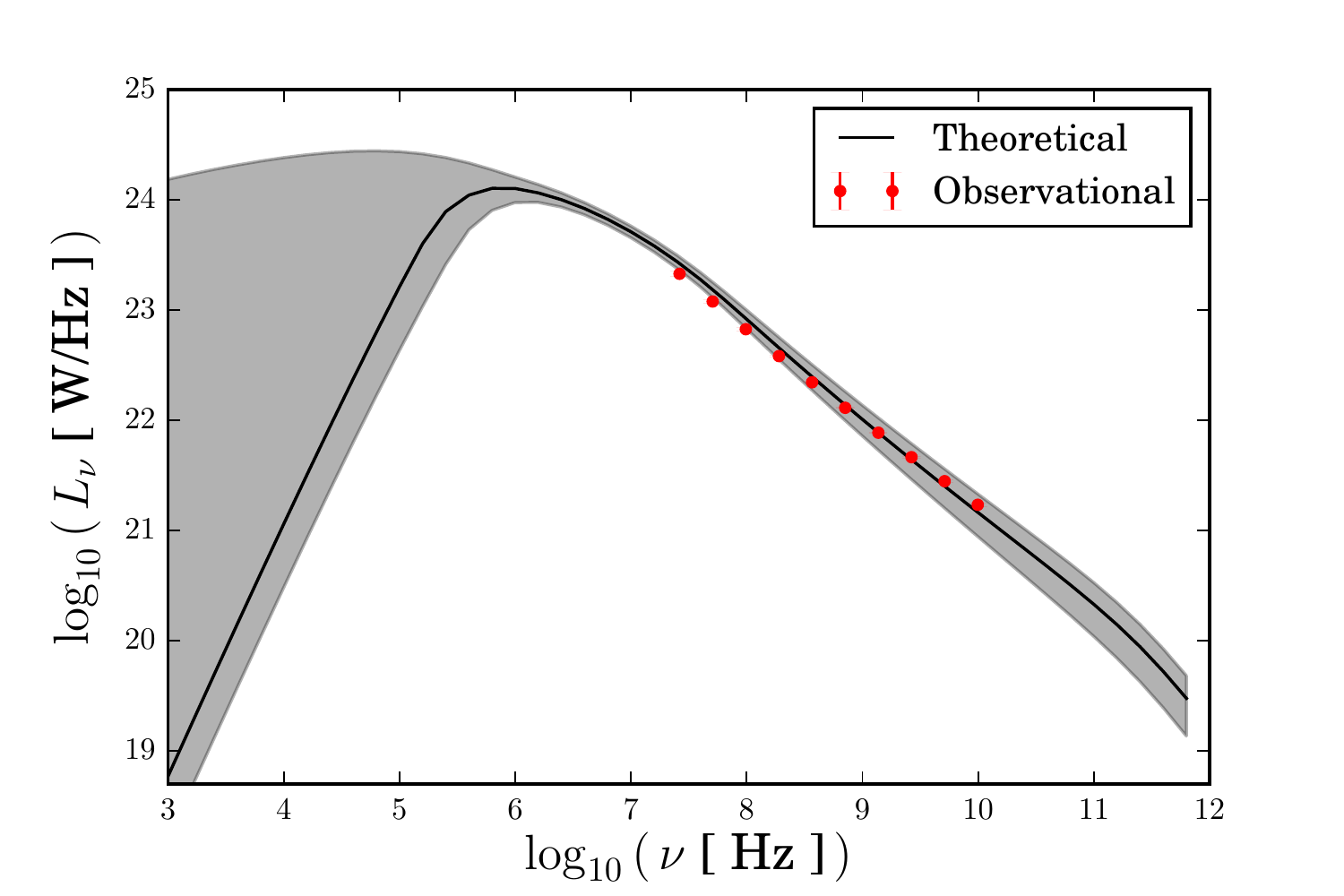}
	\caption{Comparison between the observational and theoretical luminosity spectra for the galaxy M51.  The red data points show the observational luminosity values, $L_{\nu}^{\rm obs}$.  The black line shows the theoretical luminosity $L_{\nu}^{\rm th}$ with a $1\sigma$ bound propagated from uncertainties on key physical parameters shown as a shaded region; see \secref{sec:SFGs} for details.  }
	\label{fig:SFGth}
\end{figure}

Using the measured intensity or surface brightness $S_{\nu}^{\rm obs}(r)$ of M51, its luminosity spectrum can be obtained by integrating over its area $\Omega$ on the sky,
 \begin{align}
  L_{\nu}^{\rm obs}
   &=  4\pi d^2 \int\! d\Omega \, S_{\nu}^{\rm obs}(r) ,
 \end{align}
or by integrating over its radius,
 \begin{equation}
   L_{\nu}^{\rm obs} = 4\pi \int\! dr \, 2\pi r \, C(r) \, \nu^{\alpha(r)} ,
  \label{L_freq_obs}
 \end{equation}
if we approximate M51 to be  {axi}symmetric and oriented face-on.
Values of this observed luminosity spectrum $L_{\nu}^{\rm obs}$ are shown in \figref{fig:SFGth}.

As we can only calculate $L_{\nu}^{\rm obs}$ at frequencies at which M51 has been observed, we must model a corresponding theoretical luminosity spectrum $L_{\nu}^{\rm th}$ to extrapolate to other frequencies and allow us to model a general population of SFGs.  We begin by considering the optical depth
 \begin{equation}
  \tau_{\nu} = \int_l\! \kappa_{\nu} \, dl,
 \end{equation}
found by integrating the absorption coefficient $\kappa_{\nu}$ over the path length $l$ through the absorbing/emitting medium.  At a given point on M51, its optical depth can be approximated as
 \begin{equation}
  \tau_{\nu}(r) \approx \kappa_{\nu}(r) \, l(r)
  \label{tauapprox}
 \end{equation}
where $l(r)$ is the path length along the line of sight at this point.  For a generalised SFG of inclination~$i$ and height~$h$, the path length is
 \begin{equation}
  l = \frac{h}{\cos i} .
  \label{Li}
 \end{equation}
Considering an isotropic distribution of inclination angles on the sky~\citep{Bergh1988}, the average inclination angle of an SFG is given by
 \begin{equation}
  \langle i \rangle = \int_{0}^{\pi/2} \! i \, \sin i \, di = 1\,\textnormal{rad} .
 \end{equation}

For M51, the path length has been measured to be~\citep{Berkhuijsen1997} 
 \begin{align}
  l_{\rm M51}(r) =
  \begin{cases}
   0.8\,\textnormal{kpc}  & r < 4.8\,\textnormal{kpc} \\
   1.2\,\textnormal{kpc}  & 4.8 < r < 7.2\,\textnormal{kpc} \\
   2.8\,\textnormal{kpc}  & 7.2 < r < 9.6\,\textnormal{kpc} \\
   4\,\textnormal{kpc}    & r > 9.6\,\textnormal{kpc}.  \\
  \end{cases}	
  \label{HM51}
 \end{align}
Therefore, the representative path length for an SFG of average inclination~\mbox{$\langle i \rangle$} can be expressed, assuming a constant $h$ for all SFGs, as
 \begin{equation}
  l_{\rm  SFG}
   = \frac{\cos (i_{\rm M51})}{\cos \langle i \rangle} l_{\rm M51}
   \approx 1.85 \, l_{\rm M51} \, \cos (i_{\rm M51}).
  \label{Havg}
 \end{equation}
Recalling that $\left|i_{\rm M51}\right| = 20$\,deg, \eqnrefiii{tauapprox}{HM51}{Havg}, together with the values for $\kappa_{\nu}(r)$, were used to calculate $\tau_{\nu}(r)$.

The theoretical luminosity spectrum~$L_{\nu}^{\rm th}$ is calculated as in \eqnref{L_freq_obs},
 \begin{align}
   L_{\nu}^{\rm th}
    &= 4\pi \int\! dr \, 2\pi r \, S_{\nu}^{\rm th}(r) ,
  \label{L_freq_th}
 \end{align}
with $S_{\nu}^{\rm th}(r)$ calculated using both the emission coefficient for synchrotron radiation and the absorption coefficients for both synchrotron radiation and free-free emission, per \eqnref{Sth}.  This model of the theoretical luminosity allows us to extrapolate the radio emission down to lower frequencies critical for calculating the CRB, as shown in \figref{fig:SFGth}.  To provide an estimate of the uncertainty in $S_{\nu}^{\rm th}(r)$ due to the assumed physical parameters with which we modelled M51, the errors in $\neth$, $B$ and $p$ were propagated separately into $\varepsilon_{\nu}$ and $\kappa_{\nu}$.

The radio flux of the galaxy M51 is dominated by synchrotron emission for frequencies from 27\,MHz to 23\,GHz.  Therefore, in this frequency range, free-free emission, as well as self-absorption of synchrotron emission, are considered to be negligible, implying that
 \begin{equation}
  S_{\nu}^{\rm th} \propto n_0 .
  \label{totalIsync}
 \end{equation}
Therefore, in order to determine the constant of proportionality~$n_0$ from \eqnref{ne}, we normalise $L_{\nu}^{\rm th}$ using the value of $L_{\nu}^{\rm obs}$ at a frequency of 1.4\,GHz.

\subsection{The total contribution of SFGs to the CRB}
\label{sec:SFGtot}

To estimate the total contribution to the CRB from SFGs, we consider that a typical SFG has a luminosity~$L_{\nu}^{\rm th}$ as described above.  To calculate the integrated contribution from the cosmological population of SFGs we modify the model used by~\citep{Protheroe1996} to use the 1.4\,GHz luminosity function, $\rho(L_{1.4},z)$, from~\citep{Hacking1987,Condon1992}.  This function gives the number of sources of luminosity $L_{1.4}$ at a given redshift~$z$ per unit co-moving volume per unit luminosity.  It is modelled as
 \begin{equation}
  \rho(L_{1.4},z) = g(z) \, \rho_0(L_{1.4} ,z),
  \label{rho}
 \end{equation}
where $\rho_0(L_{1.4} ,z)$ is the local luminosity function (at $z=0$), and $g(z)$ is the density evolution function.   \citep{Protheroe1996} gives
\begin{equation}
    g(z) = 1
\end{equation}
and
 \begin{equation}
  \rho_0(L_{1.4},z) = 2.94 \times 10^{28} \, 10^Q \, \textnormal{Mpc$^{-3}$\,(W\,Hz$^{-1}$)$^{-1}$},
 \end{equation}
where~\citep{Condon2002}
 \begin{equation}
  \begin{aligned}
   Q = Y
    &- \bigg[ B^2+\bigg( \frac{\log_{10}[L_{1.4} / f(z)] - X}{W} \bigg)^{\!\!2} \bigg]^{\!1/2} \\
    &- 2.5\log_{10}[L_{1.4} / f(z)]
   \label{Q}
  \end{aligned}
 \end{equation}
with $L_{1.4}$ in W\,Hz$^{-1}$ and fitted parameters from \tabref{tab:cosmoparams}.  Furthermore, in Eq.~\ref{Q} above, $f(z)$ is the luminosity evolution function, given in~\citep{Protheroe1996} by
 \begin{equation}
  f(z) = 
   \begin{cases}
    (1+z)^4    & z < z_0 \\
    (1+z_0)^4  & z \geq z_0 \\
   \end{cases}	
  \label{fz}
 \end{equation}
where $z_0 = 1.2$.

\begin{table}
 \centering
 \begin{threeparttable}
  \caption{Fitted parameters from \citet{Condon2002} for evolutionary model in \eqnref{Q}.}
  \begin{tabular}{lrrrr}
   \toprule
   Source type & \chead{B} & \chead{W} & \chead{X}    & \chead{Y}    \\
   \midrule
   SFGs        &       1.9 &      0.67 &     22.35    &      3.06    \\
   RGs         &       2.4 &      0.78 &     25.8\phn &      5.6\phn \\
   \bottomrule
  \end{tabular}
  \label{tab:cosmoparams}
 \end{threeparttable}
\end{table}

Following \citet{Protheroe1996}, the total intensity from all SFG sources can be expressed as
 \begin{equation}
  \begin{aligned}
   I_{\nu} = \frac{1}{4\pi}
     &\int_{0}^{z_r}\! dz \, \frac{dV_c}{dz} \, \frac{1+z}{4\pi d_L {(z)}^2} \\
    \times
     &\int_{0}^{\infty} dL_{1.4} \, \rho(L_{1.4},z) \, \frac{L_{\nu'}}{ {L_{\nu' ,1.4}}} \, L_{1.4}
   \label{Iztotal}
  \end{aligned}
 \end{equation}
where the co-moving volume is defined as (e.g.~\citep{Hogg1999})
 \begin{equation}
  \frac{dV_c}{dz} = d_H \, \frac{d_L {(z)}^2}{E(z)}.
  \label{Vc}
 \end{equation}
In this equation, $d_L {(z)}$ is the luminosity distance and $d_H = c/H_0$ is the Hubble distance~\citep{Peebles}.  The function $E(z)$ is defined as
 \begin{equation}
  E(z) = \sqrt{\Omega_M \, (1+z)^3 + \Omega_k \, (1+z)^2 + \Omega_{\Lambda}},
  \label{E(z)}
 \end{equation}
where the density parameters $\Omega_M$, $\Omega_k$ and $\Omega_{\Lambda}$ are as defined in \secref{sec:intro}.

In \eqnref{Iztotal} above, the notation $L_{\nu'}$ refers to the luminosity in the source reference frame at frequency $\nu'=\nu (1+z)$, which is Doppler shifted to the observed frequency $\nu$ by the expansion of the Universe.   {Furthermore, $L_{\nu',1.4}$ refers to the value of $L_{\nu'}$ at $1.4~\textnormal{GHz}$, which was used for normalising.}

The factor of $1+z$ in \eqnref{Iztotal} is a combined effect of the relation between~$d_L {(z)}$ and the  {comoving} distance at the time of emission, $d=d_L {(z)}\,(1+z)^{-1}$, and the Doppler shift, since the intensity~$I_{\nu}$ is defined at the observation frequency.  We set the upper limit of the integral in \eqnref{Iztotal} to be $z_r \sim 10$.  The effect of this limit is discussed further in \secref{Sec:redshiftlimit}.

\section{Radio galaxies}
\label{sec:RGs}

Radio galaxies (RGs) are some of the most luminous sources in the radio sky.  They consist of a parent galaxy hosting an active galactic nucleus (AGN) generating symmetric radio jets, which terminate in lobes as they interact with the extragalactic medium.  The lobes of an RG contain magnetised relativistic plasma which produce radio emission through synchrotron radiation~\citep{DeYoung1976}.

RGs are divided into two Fanaroff-Riley classes, FR\,I and FR\,II, depending on their emission and morphological properties~\citep{FR1974}.  FR\,I sources tend to be less luminous, with their radio emission dominated by their jets and central AGN, while FR\,II sources are dominated by their two lobes, typically being observed in radio as a double source.  FR\,II sources are also more numerous.  For example, the Combined NVSS and FIRST Galaxies catalogs (CoNFIG; \citep{config1,config2}) contains 859~resolved sources selected from the NVSS survey~\citep{nvss} within the northern field of the FIRST survey~\citep{first}.  Almost all ($>95$\%) of these sources are RGs that have been morphologically classified into a Fanaroff-Riley class, and of these, approximately 90\% are FR\,II RGs.

\begin{figure} 
	\centering
	\includegraphics[scale=0.40]{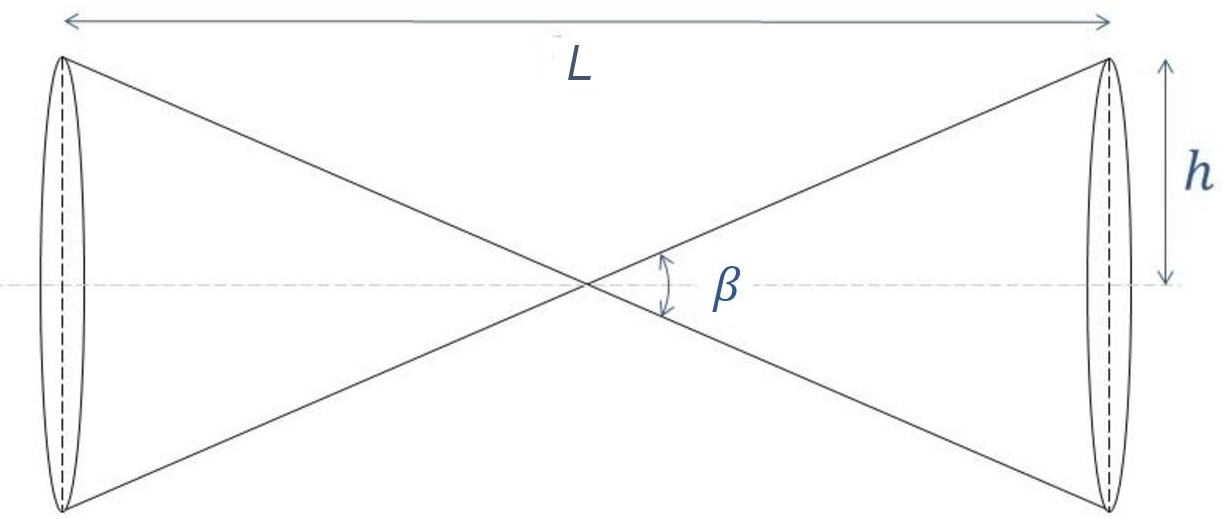}
	\caption{The simplified geometry used to describe RGs.  As described in the text, $L$ is the physical size, $h$ the maximum radius and $\beta$ the opening angle.  This geometry is assumed to be perpendicular to the observer's line of sight.}
	\label{fig:geometry}
\end{figure}

As FR\,II sources dominate the contribution of RGs to the CRB, we calculate this contribution with an appropriately lobe-dominated model, using the double-cone geometry shown in \figref{fig:geometry}.  We approximate the overall geometrical shape of a typical RG as two symmetric conical lobes, seen at an inclination of $\sim0\,$deg with a projected opening angle of $\beta = 22.6\,$deg, based on average values found for such sources~\citep{Pushkarev2017}.  The average path length through this system in the observed direction is
 \begin{equation}
  l = \frac{\pi h}{4} 
 \end{equation}
where $h$ is the maximum radius of the cone as seen by the observer, illustrated in \figref{fig:geometry} and calculated following
 \begin{equation}
  h = \frac{L}{2} \tan \frac{\beta}{2} .
 \end{equation}
 
\subsection{The typical luminosity spectrum of RGs}

We use the 3CRR survey~\citep{3CRR_Survey} at 178\,MHz to estimate a representative RG spectrum, under the assumption that each galaxy in the survey can be described by the geometry illustrated in \figref{fig:geometry}.  The spectra for 171 of the 173 galaxies in the 3CRR survey were simulated as described in the following paragraphs before their derived luminosity spectra were averaged to give a representative RG spectrum, $L_{\nu'}$.  The remaining two galaxies were excluded as they had no recorded angular or linear size.  A discussion of their exclusion, as well as a more detailed description of the survey can be found in \secref{Dis:survey Comp}.

Following \citet{Protheroe1996} we define the electron energy density distribution for an RG as
 \begin{equation}
  n_{\rm e}(E) = n_0
   \begin{cases}
    (E/E_0)^p  & E \geq E_0 \\
    0          & E < E_0 \\
   \end{cases}
  \label{RG_energy_density}
 \end{equation}
with an amplitude~$n_0$, an energy spectral index~$p$ and a minimum cut-off energy $E_0=100$\,MeV~\citep{Condon1984}.  The value of this cut-off is  {roughly} a consequence of the electron energy being limited by the mass of the pion, as the relativistic electron population is  assumed to be produced by (indirect) pion decay~\citep{Protheroe1982}.  {See~\secref{Sec:disc:cut} for a more detailed discussion of this low-energy cut-off.}

Where \citet{Protheroe1996} assume that all RGs have a spectral index of $\alpha=-0.75$ and consequently an energy density spectral index of $p=2.5$, we use the specific spectral indices for each individual galaxy in the 3CRR survey.  We estimate the values of $n_0$ and magnetic field strength~$B$ for each galaxy by assuming equipartition, as is common in situations where magnetic fields are difficult to measure directly.  We assume that the total energy density in a galaxy can be expressed as~\citep{Pacho}
 \begin{equation}
  u_{\rm tot} = u_B + u_{\rm e},
  \label{Utot}
 \end{equation} 
where $u_B$ is the magnetic field energy density given by
 \begin{equation}
  u_B = \frac{B^2}{2\mu_0}
  \label{UB}
 \end{equation}
and $u_{\rm e}$ is the electron energy density given by
 \begin{equation}
  u_e = n_0 \int_{0}^{\infty}\! E \, \frac{n_{\rm e}(E)}{n_0} \, dE
  \label{Ue}
 \end{equation}
where $\mu_0$ is the permeability of free space.  Equipartition assumes that, for a system with interacting particles, $u_e\sim u_B$, thus providing a relationship between the parameters $n_0$ and $B$.

Under the assumption of equipartition, the synchrotron intensity at 178\,MHz was modelled for each galaxy, similarly to the procedure described in \secref{sec:SFGs}, but using \eqnref{RG_Intensity} instead of \eqnref{Sth}, omitting free-free absorption due to the lack of thermal electrons.  Using the observed intensities, we calculated the equipartition magnetic field $B_{\rm eq}$ and the corresponding value of $n_0$ for each galaxy.

\begin{figure}
    \centering
    \includegraphics[width=\linewidth]{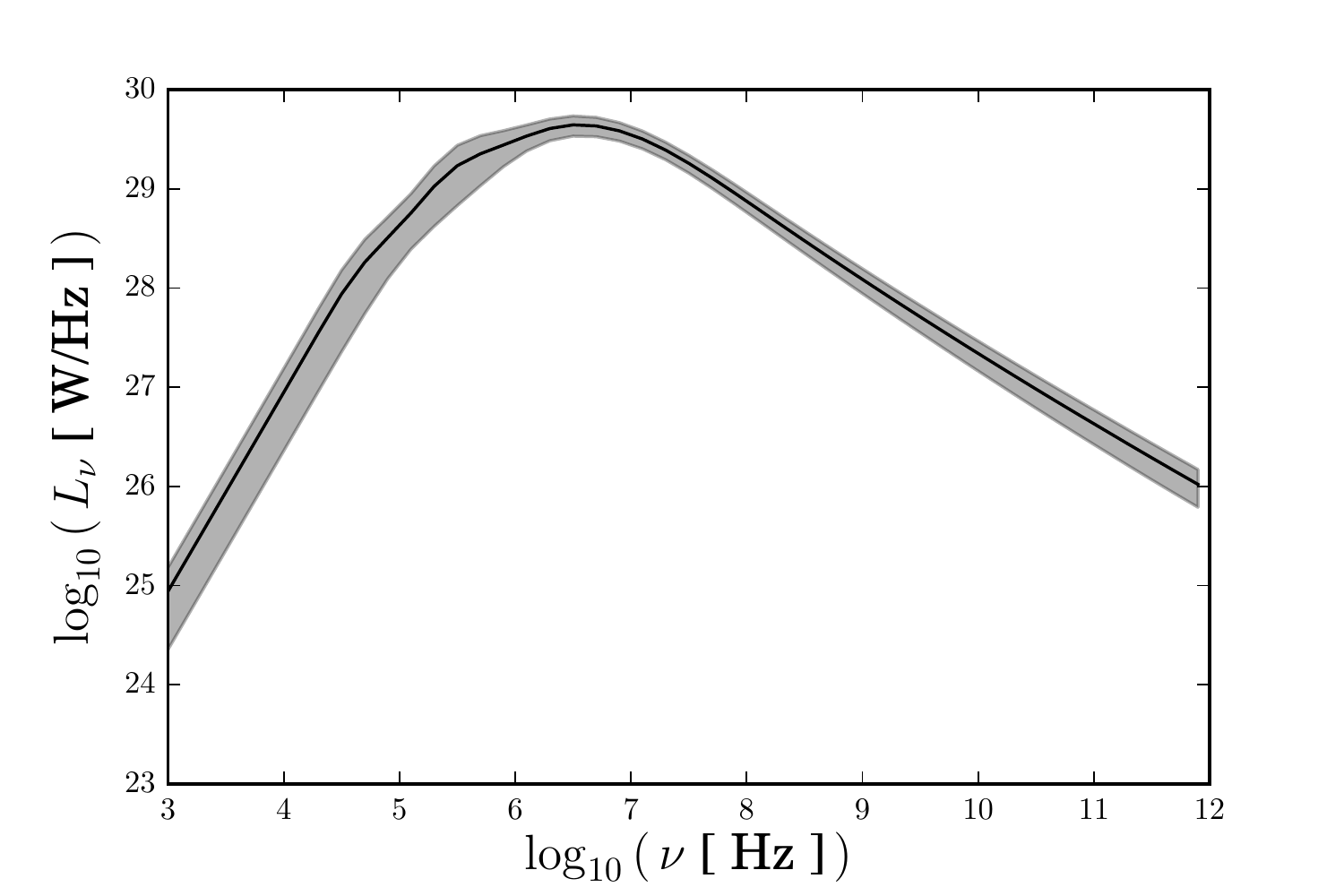}
    \caption{The representative RG spectrum used to derive the contribution of RGs to the CRB; see \secref{sec:RGs} for details.}
    \label{fig:Typical_RG}
\end{figure}

The spectral flux density~$F_{\nu}$ was then determined for each galaxy as well as the luminosity in the reference frame of the galaxy
 \begin{equation}
   {L_{\nu'}} = \frac{ 4 \pi d_L {(z)}^2 }{ 1 + z } F_{\nu},
 \end{equation}
where $d_L {(z)}$ is the luminosity distance to each galaxy.  The average over all galaxy luminosities is then used to provide the representative luminosity spectrum for RGs, $L_{\nu'}$, illustrated in \figref{fig:Typical_RG}.

\subsection{The total contribution of RGs to the CRB}

To extrapolate from the emission of a typical RG to the total contribution of such sources to the CRB, we use the same method as described for SFGs in \secref{sec:SFGtot}.  The same evolution model as in \eqnref{Q} was used, but with the appropriate parameters for radio galaxies from \tabref{tab:cosmoparams}, $g(z) = 1$, and $z_0 = 0.8$.  We note that the previous \citet{Protheroe1996} estimate used a parameterisation from \citet{Condon1984}, based on a now-outdated cosmology with a Hubble constant of $H_0 = 50$\,km\,s$^{-1}$\,Mpc$^{-1}$.

\section{Comparison to observations and previous estimates}
\label{sec:comparison}

To validate our estimate of the CRB we compare it to observational source-count data and corresponding models from~\citet{Massardi2010}, \citet{deZotti2010} and \citet{Bonato2017}.  These are available at a number of different frequencies towards the high end of our modelled frequency range, and provide separate data for the contributions of SFGs and RGs.  To calculate the total intensity from each of these models we integrate the source-count models using a lower limit consistent with the observational threshold in each case.

We also compare our estimate with the previous estimate of the CRB by \citet{Protheroe1996}, which was also separated into contributions from SFGs and RGs.  \citeauthor{Protheroe1996} provided estimates under two different assumptions regarding the cosmological evolution of their sources.  In one case, they used a luminosity function based on the evolution of infrared sources as a proxy for the evolution of radio sources, assuming radio and infrared flux to be well correlated.  In the other, due to uncertainty regarding this correlation for low-luminosity sources (see their Fig.~4), they considered radio sources not to evolve over cosmological time.  In this work, we use a more recent luminosity function defined directly in terms of radio flux (see \eqnref{Q}), so we are not subject to this uncertainty, and present results only for the case in which evolution of sources does occur.  We therefore, in this section, compare our result with the corresponding result from \citeauthor{Protheroe1996}, which also includes the effects of evolution.

\subsection{Contribution from SFGs}

\begin{figure}
	\centering
	\includegraphics[width=\linewidth]{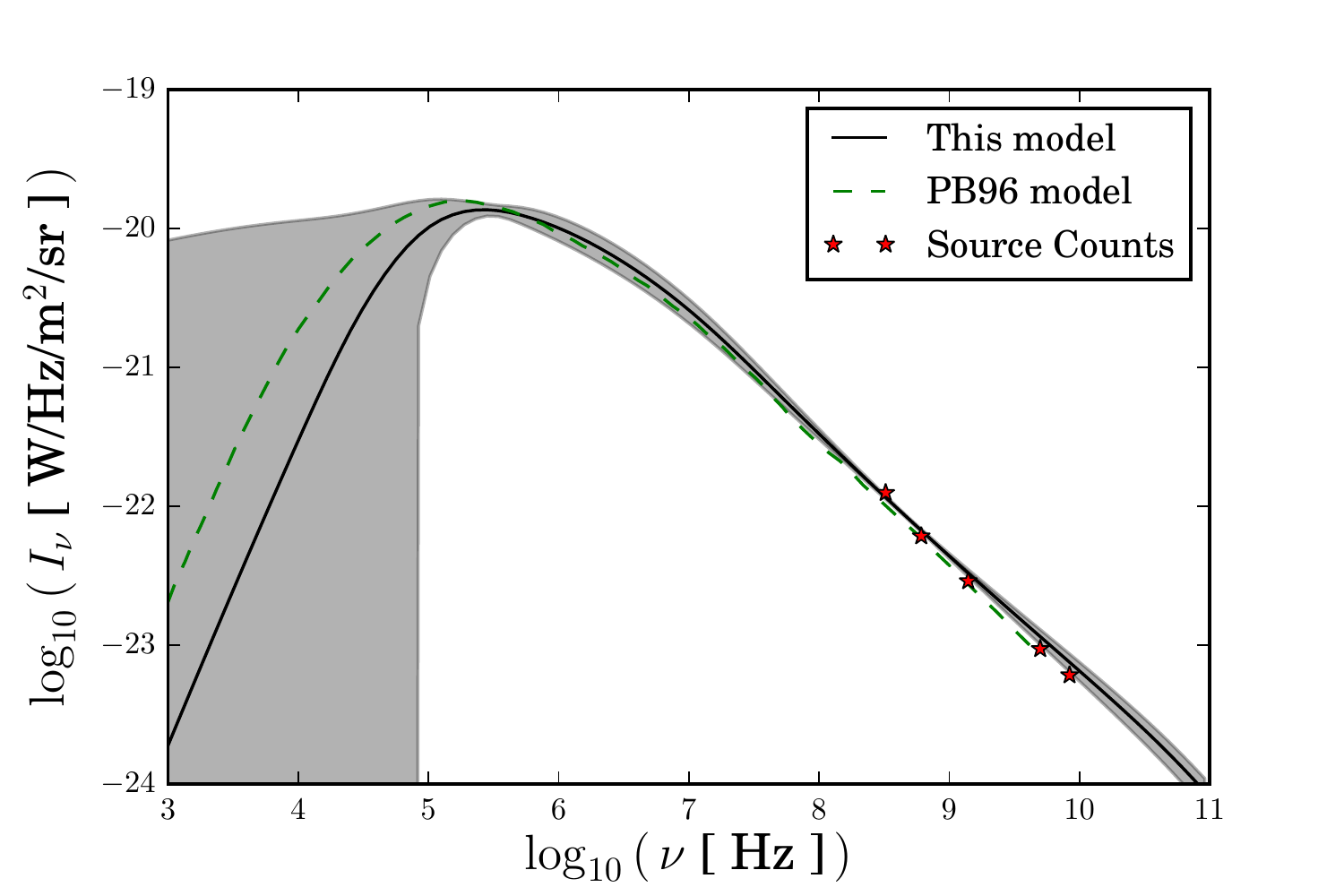}
	\caption {The contribution of SFGs to the CRB as calculated here (solid black line) and by \citet{Protheroe1996} (their Fig.~5; dashed green line).  Observational measurements from radio source-count data~\citep{Massardi2010,deZotti2010,Bonato2017} are shown as stars at discrete frequencies.}
	\label{fig:SFG_Background}
\end{figure}

In \figref{fig:SFG_Background} we compare the estimate of the SFG contribution to the CRB calculated in this work to that from \citet{Protheroe1996}, as well as to the empirical values calculated from the source-count models for SFGs alone.  Both estimates are generally consistent with the values calculated from the source counts.  We note that the source-count data also have associated uncertainties, however these are too small to be seen in the plot.

At lower frequencies, we observe an offset between the estimate from this work and that of \citet{Protheroe1996}.  Although both estimates are consistent within the $1\sigma$ uncertainty, there is a notable difference in the position of the peak in the spectrum.  This difference is a consequence of our differing electron density and temperature assumptions.  A detailed discussion of how these parameters affect the shape of the spectrum can be found in \secref{sec:disc:vare}.

\subsection{Contribution from RGs}

\begin{figure}
    \centering
    \includegraphics[width=\linewidth]{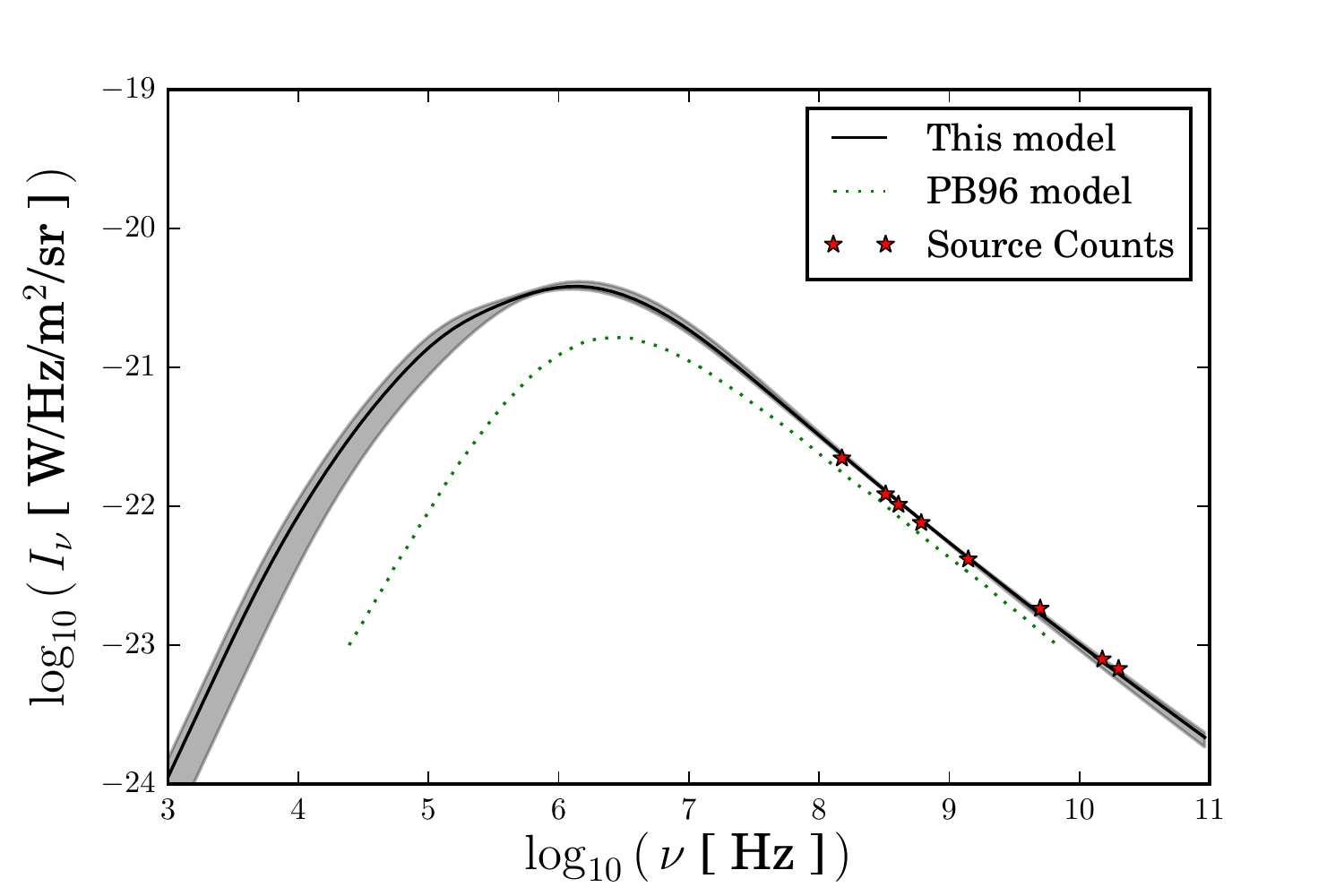}
    \caption{The contribution of RGs to the CRB as calculated here (solid black line) and by \citet{Protheroe1996} (dashed green line).  Observational measurements from radio source-count data~\citep{Massardi2010,deZotti2010,Bonato2017} are shown as stars at discrete frequencies.}
    \label{fig:RG_Background}
\end{figure}

\begin{figure}
    \centering
    \includegraphics[width=\linewidth]{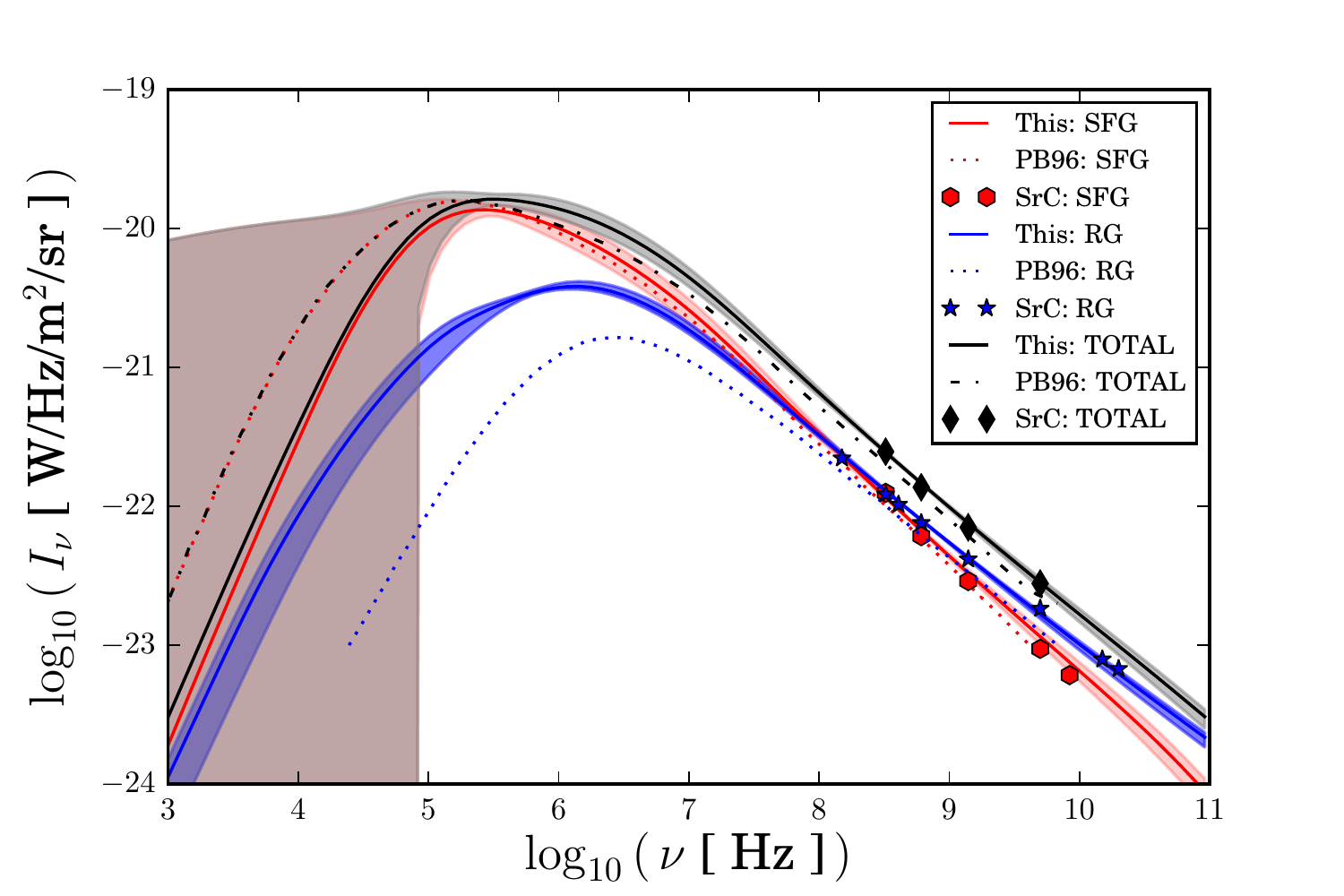}
    \caption{The total CRB is shown along with the contributions from SFGs and RGs as calculated in this work (solid lines) and by \citet{Protheroe1996} (dashed \& dotted lines).  Observational measurements from radio source-count data~\citep{Massardi2010,deZotti2010,Bonato2017} are shown as stars, diamonds and hexagons at discrete frequencies.}
    \label{fig:TotalBackground}
\end{figure}

In \figref{fig:RG_Background} we compare the estimate of the RG contribution to the CRB to that of \citet{Protheroe1996}.  In this case there is a clear difference between the two estimates.  While the estimate from this work agrees with the values from the radio source-count data, the model of~\citep{Protheroe1996} underestimates these values significantly at all frequencies.  This discrepancy is also evident in the total CRB estimate, see \figref{fig:TotalBackground}.

A notable difference between the estimate from this work and that of \citet{Protheroe1996} at higher frequencies is the slight curvature in the spectral shape produced by this work.  This is due to the use of multiple RG spectra with different spectral indices, which allows for a non-linear spectrum  {in logarithmic space} at high radio frequencies.  This distribution of spectral indices is of particular importance when considering populations of galaxies across a wide range of redshifts where inverse-Compton losses may cause steepening of spectra at higher redshifts, as has been observed previously for 3CR~\citep{morabito2018}.

At lower radio frequencies, as well as predicting generally larger values of the intensity, our model has a different slope to the original \citet{Protheroe1996} model.  This property is a consequence of using an exact treatment for the synchrotron self-absorption process in each RG rather than a linear approximation.  For a more detailed description of this see \secref{sec:disc:SSA}.

\section{Discussion of uncertainties}
\label{sec:disc}

\subsection{The contribution of SFGs} 
\label{sec:disc_SFG}

 {The new Voyager 2 observations of our galaxy~\citep{stone2019} constitute a unique insight into the electron energy density of an SFG.  These measurements agree with the model used in this analysis, as per Eq.~\ref{ne}, at all but very low energies.  However, we cannot determine whether this is a general result for SFGs or is particular to the Very Local Interstellar Medium of our galaxy observed by Voyager 2.  Further, in this analysis physical properties such as the magnetic field, spectral indices and observed luminosity at $1.4~$GHz specific to our average galaxy, M51, were used to create the average luminosity spectrum.  The same information is not available for our own galaxy.  Consequently, use of the Voyager electron energy density would lead to inconsistencies in our model.  For these reasons, we chose to use the power-law model presented in Eq.~\ref{ne} for estimating the electron energy density of SFGs.}

The largest source of uncertainty amongst those discussed in \secref{sec:SFGs} when calculating the contribution of SFGs to the CRB arises from the loose observational constraints on the thermal electron density.  It can also be observed from \eqnrefii{kap_f}{eps_f} that the free-free emission and absorption are stronger functions of the electron density than of the electron temperature.  

\begin{figure*}
	\centering
	\includegraphics[width=.49\linewidth]{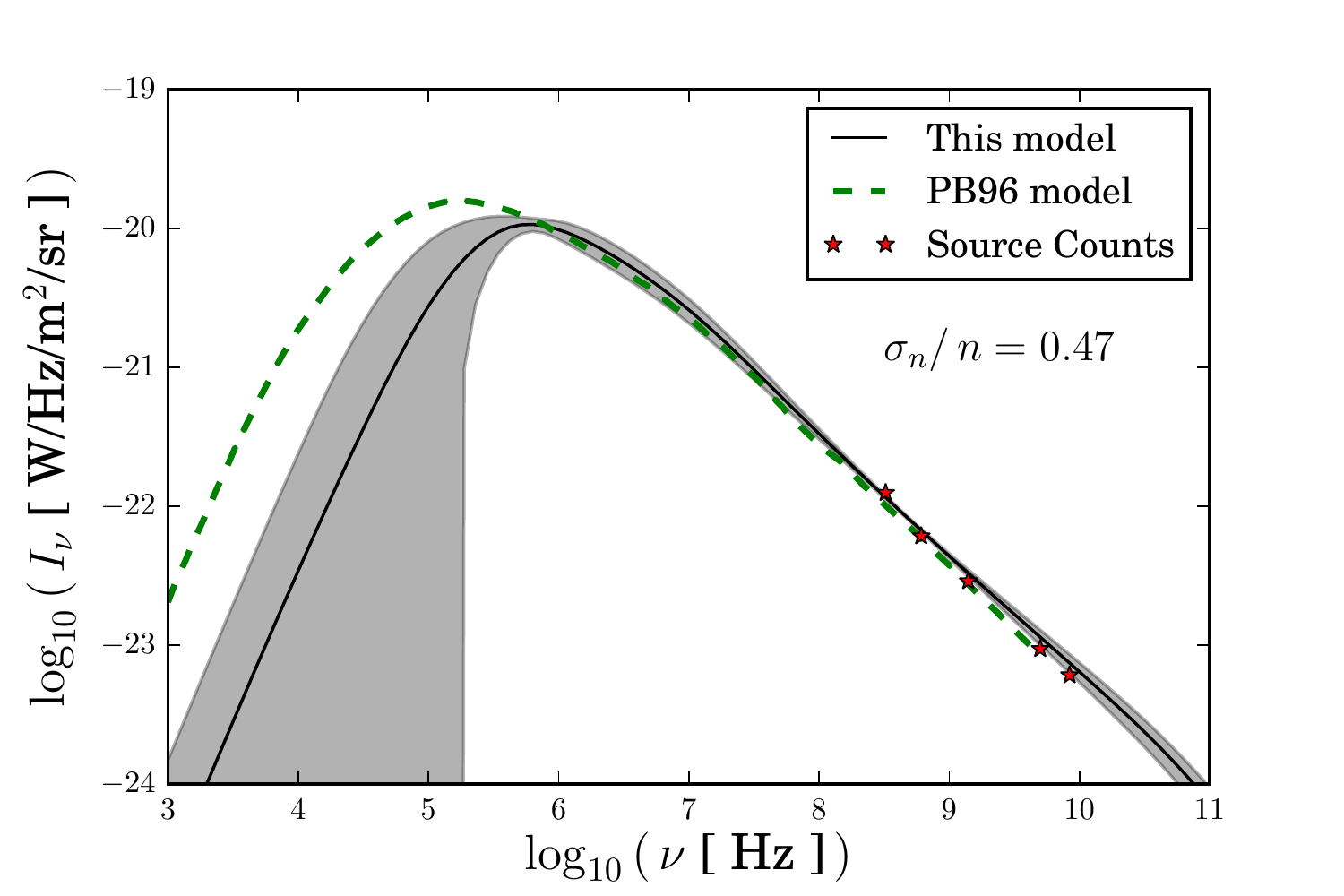}
	\includegraphics[width=.49\linewidth]{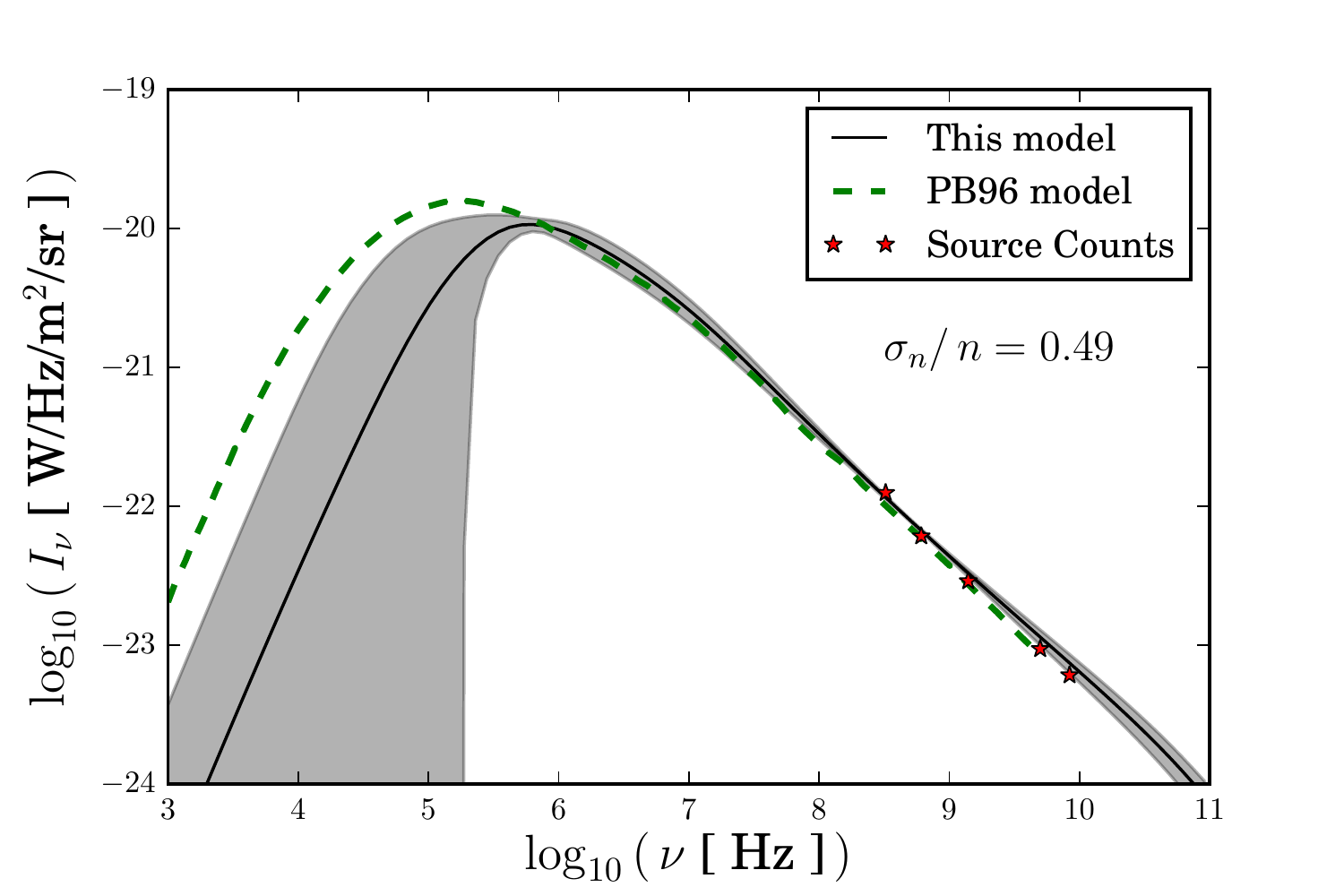} \\
	\includegraphics[width=.49\linewidth]{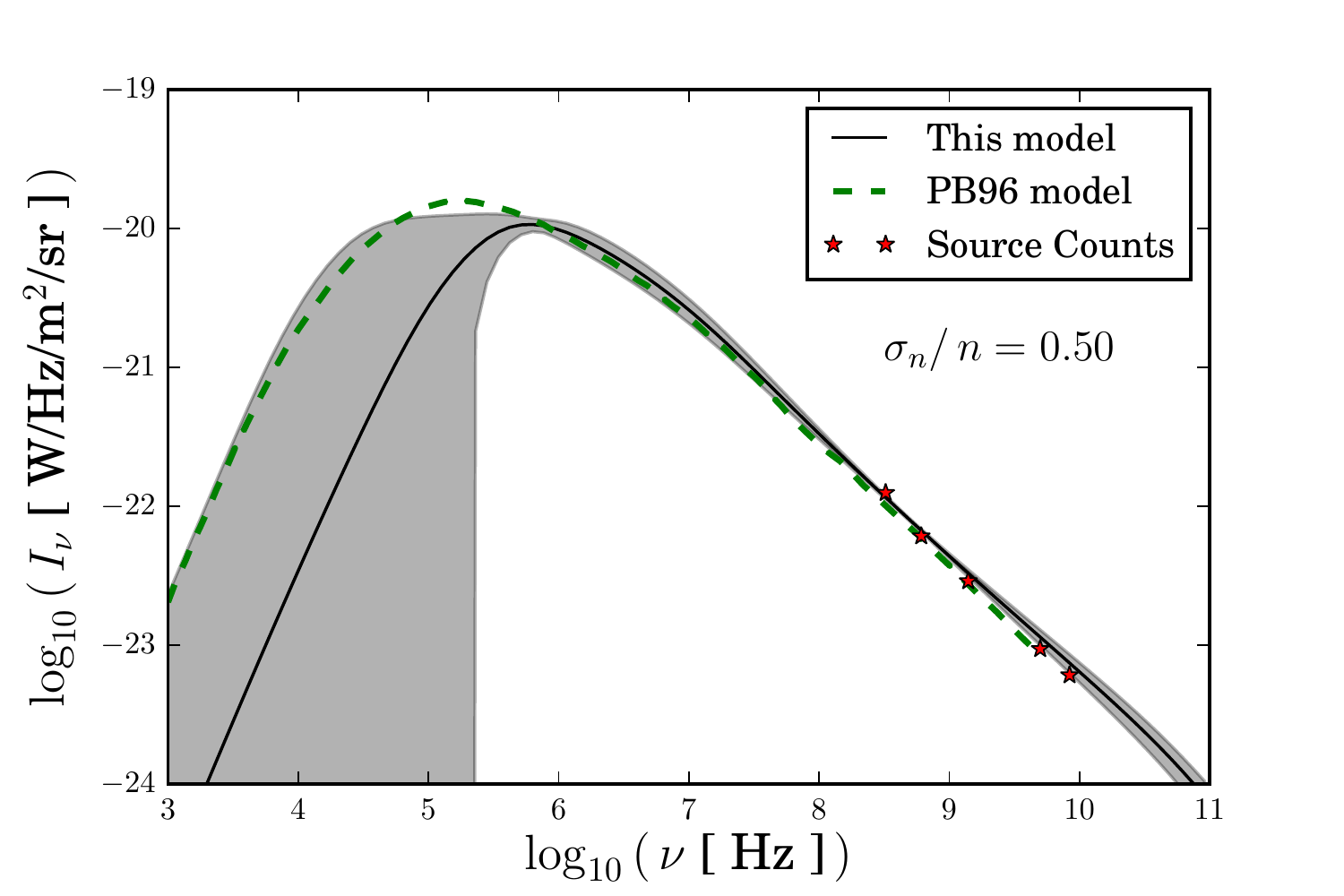}
	\includegraphics[width=.49\linewidth]{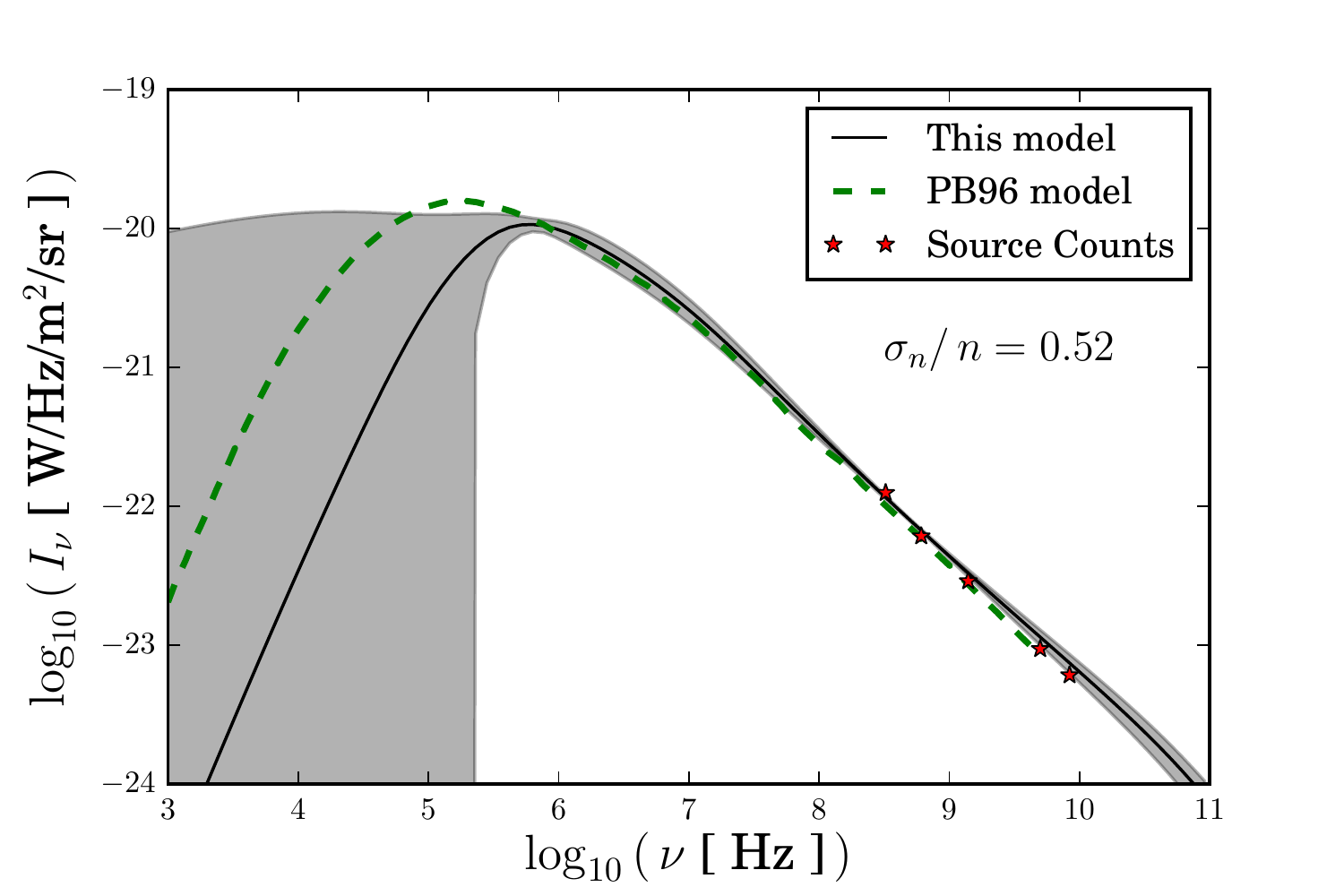}
	\caption{The variation of the uncertainty in the CRB from SFGs with the error on the electron density $\neth$.  For these calculations a value of $n_{e,th}=0.02$\,{cm}$^{-3}$ was used, equivalent to the radially-weighted average of the values used for the SFG model in \secref{sec:SFGs}; see \secref{sec:disc_SFG} for details.}
	\label{fig:errn}
\end{figure*}

The large error in $\neth$, which ranges from $\sim 16$ to~200\%, propagates through to the SFG CRB estimate predominantly at low frequencies where its effect can be seen as a rapid increase in the uncertainty in the CRB estimate.  By considering artificially accurate measurements of $\neth$ with uncertainties in the range 0--100\%, we find that a precision of $\sigma_{\neth}/\neth \sim 50$\% (see \figref{fig:errn}) would allow the estimated CRB from this work to be distinguished from that of \citet{Protheroe1996} at $1\sigma$ confidence.

\subsubsection{Effect of varying $\neth$ and $T_{\rm e}$}
\label{sec:disc:vare}

\begin{figure}
	\centering
	\includegraphics[width=\linewidth]{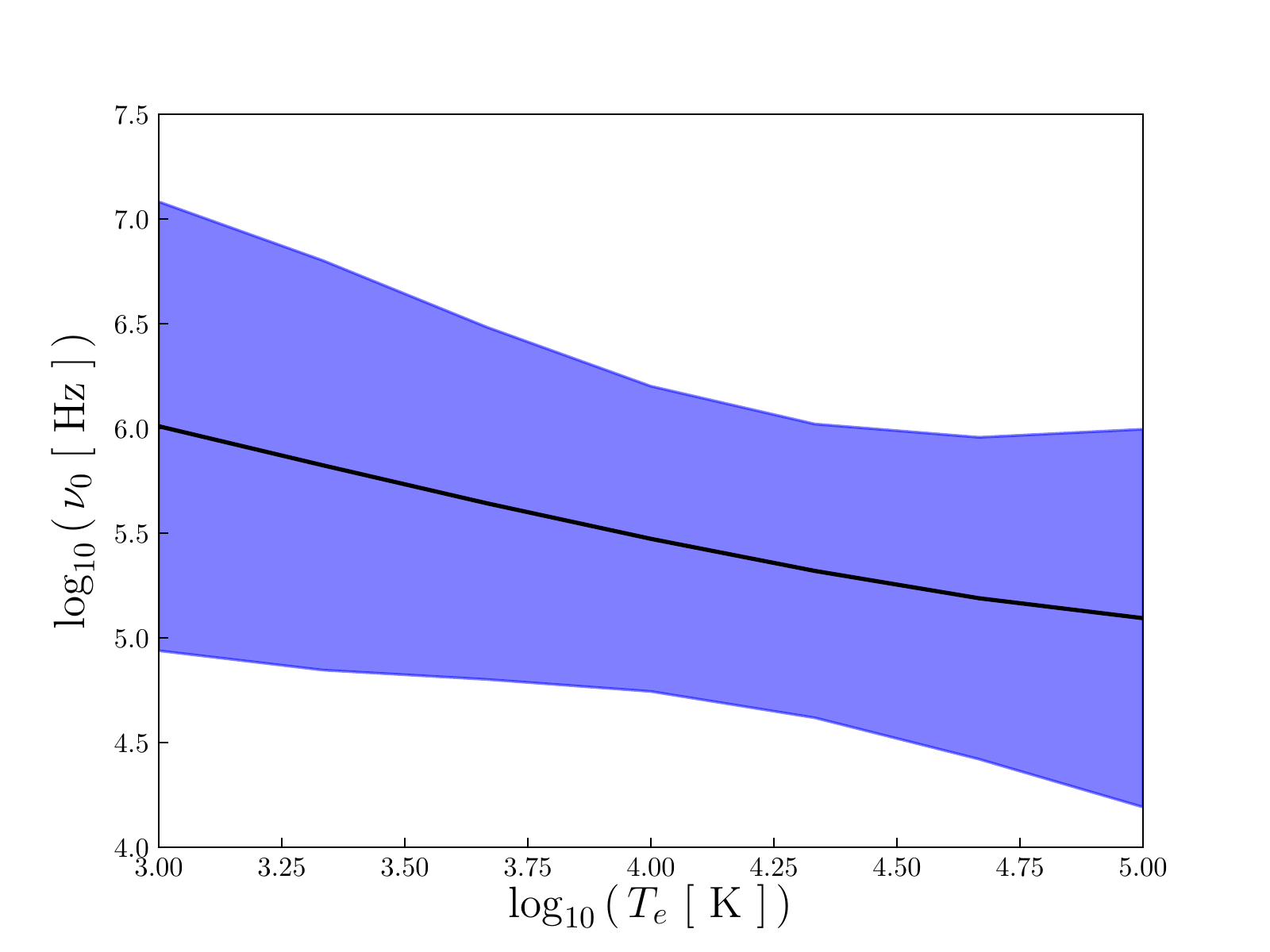}
	\includegraphics[width=\linewidth]{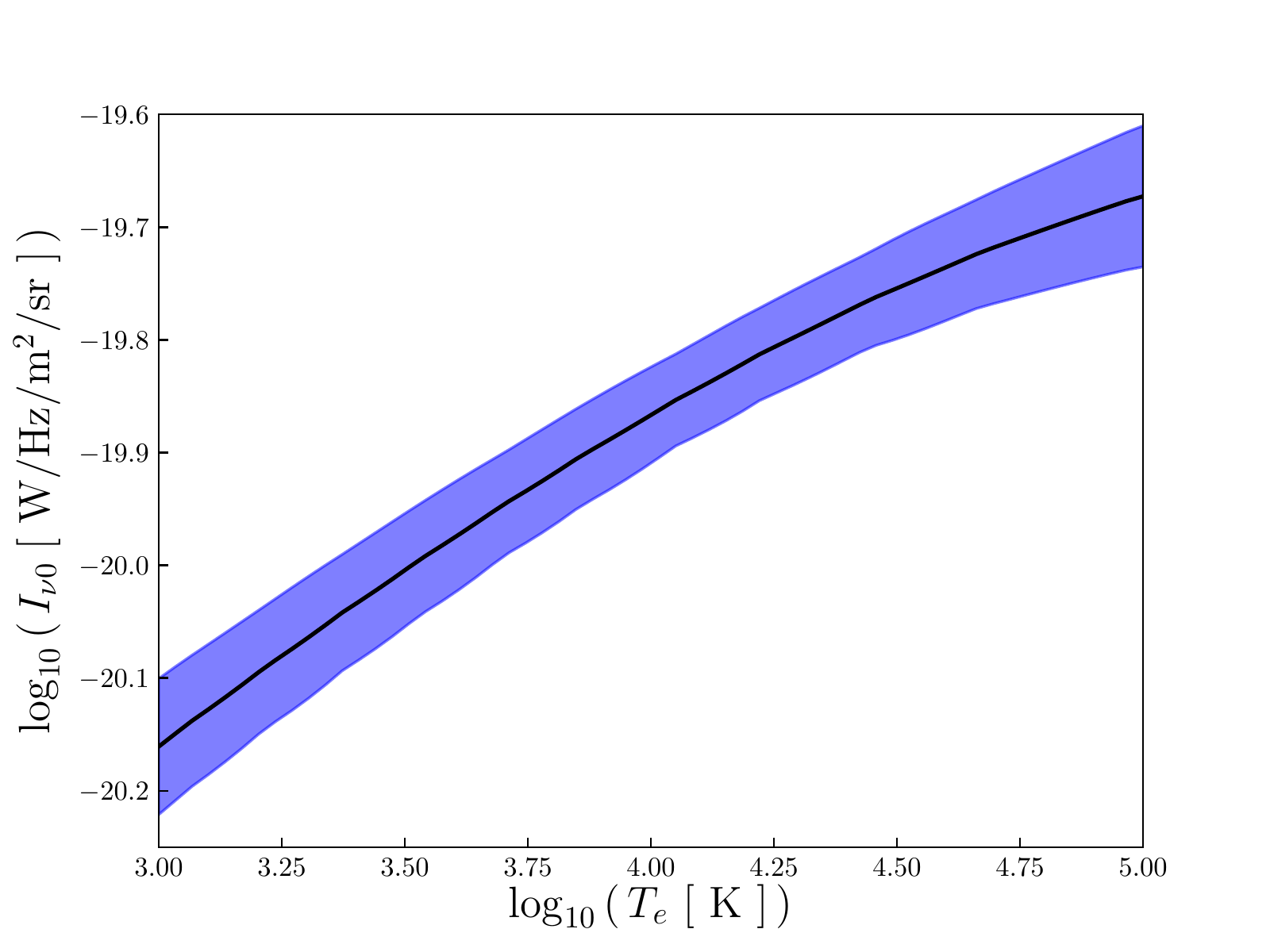}
    \caption{Top: variation of peak frequency~$\nu_0$ with electron temperature~$T_{\rm e}$ for SFGs.  Bottom: equivalent variation of the peak intensity $I_0$.  Shaded areas in both plots show $1\sigma$ uncertainties.}
    \label{fig:Tevs}
\end{figure}

In \secref{sec:SFGs}, we assumed perfect knowledge of the electron temperature~$T_{\rm e}$.  In order to understand the effect of changing this parameter we now vary the value of $T_{\rm e}$ whilst maintaining the \citet{Berkhuijsen1997} values for $\neth$.  The results of this test are shown in \figref{fig:Tevs}.
	
We observe that as the temperature increases the free-free absorption in our representative SFG spectrum greatly decreases.  This is expected because $\kappa_{\nu}^F \propto T_{\rm e}^{-1.35}$.  As a consequence the peak frequency, $\nu_0$, decreases and the peak intensity, $I_{\nu,0}$, increases, as can be seen in \figref{fig:Tevs}.  There is also a decrease in free-free emission, since $\varepsilon_{\nu}^F \propto T_{\rm e}^{-0.5}$, but this is less significant as the dependence on temperature is weaker and the free-free emission is subdominant to the synchrotron intensity.  

The value of the peak frequency is defined as the point where the spectral index is zero.  In order to capture the asymmetry in the CRB intensity spectrum we calculate approximate bounds on this position using the error on the spectral index, $\sigma_{\alpha = 0}$, and take the bounding frequencies equivalent to the point in the spectrum where the tangent to the spectral index  at $\alpha = 0$ intercepts the lines of constant $\pm\sigma_{\alpha  = 0}$, see \figref{fig:Tevs}.
	
\citet{Protheroe1996} assumed a comparatively high value of $T_{\rm e} = 3 \times 10^5\,$\,K, more akin to the hot ISM than the warm ISM responsible for free-free absorption.  The result of this is that the peak in their spectrum is pushed to lower frequencies than that calculated in this work, as can be seen in \figref{fig:SFG_Background}.

\begin{figure}
	\centering
	\includegraphics[width=\linewidth]{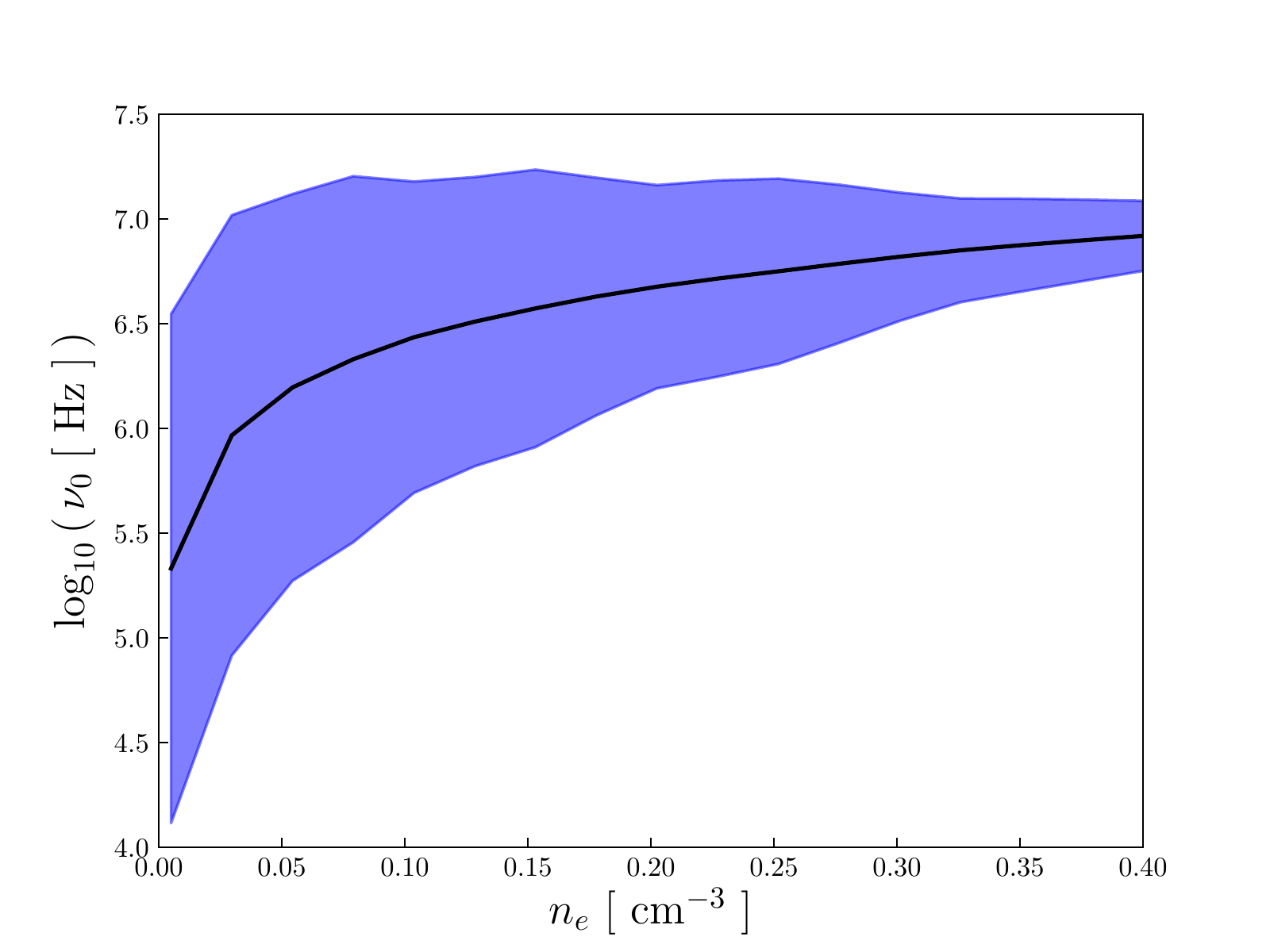}
	\includegraphics[width=\linewidth]{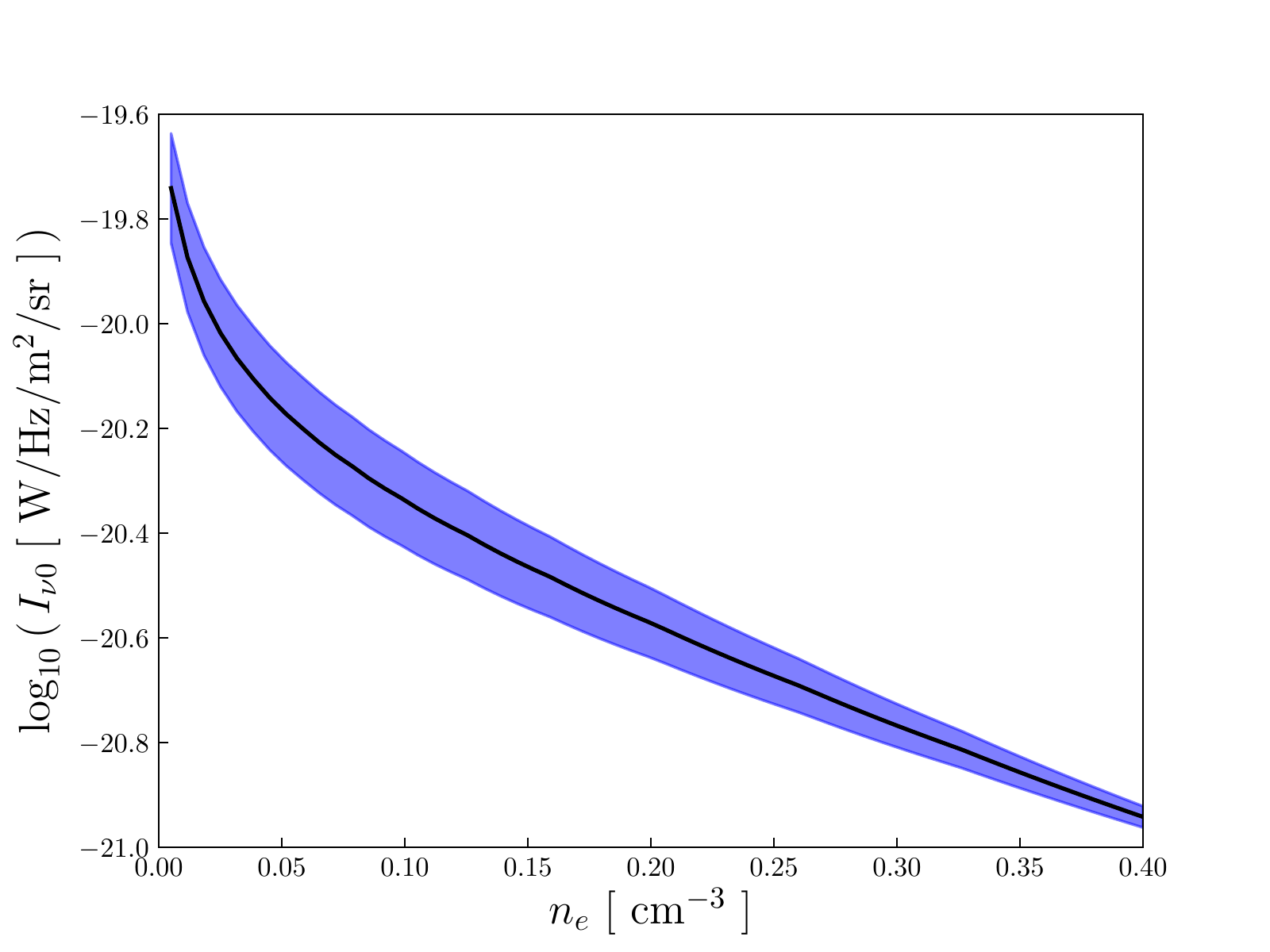}
	\caption{Top: variation of peak frequency~$\nu_0$ with electron density~$\neth$ for SFGs.  Bottom: equivalent variation of the peak intensity~$I_0$.  Shaded areas in both plots show $1\sigma$ uncertainties.}
	\label{fig:nevs}
\end{figure}

Whilst holding the electron temperature constant at $T_{\rm e}=10^4$\,K we also consider the effect of varying $\neth$.  In order to provide a clear understanding of the changes introduced by this test we assume here absolute precision in $\neth$ and a uniform electron density across the SFG.  The results of this test are shown in \figref{fig:nevs}.  It can be seen that as $\neth$ increases, the peak frequency, $\nu_0$, also increases, and the peak amplitude, $I_{\nu,0}$, decreases.  This is because $\kappa_{\nu}^F \propto \neth^{2}$ and so as $\neth$ increases so does the free-free absorption, resulting in a decreased peak intensity.  Again there is an associated effect on the free-free emission, but as before free-free emission is not a dominant factor in determining the emission from an SFG and so this is less significant than the effects on $\kappa_{\nu}^F$.
	
The variation of $\nu_0$ and $I_{\nu,0}$ with $\neth$ can be seen in \figref{fig:nevs}.  \citet{Protheroe1996} used a value of $n_{e,th}=0.01$\,{cm}$^{-3}$, which is smaller than the radially-weighted average of \eqnref{BerkElecDen} and leads to the comparatively higher peak amplitude that can be seen in (e.g.) \figref{fig:errn}.

\subsection{The contribution of RGs}
\label{Dis:survey Comp}

The 3CRR survey, used here to estimate the contribution of RGs to the CRB, is built from the 3CR, 4C and 4CT surveys~\citep{3cr-survey,4c-survey,4ct-survey}.  This was done with an aim of including galaxies that were missed in the original 3C survey due to observational limitations but that otherwise meet the criteria for survey completeness.

The survey is estimated to be complete to $96\,\%$ for sources with $\theta < 10\,$arcmin and $F \geq 10\,$Jy at 178\,MHz in the region $\delta \geq 10\,$deg and $|b| \geq 10\,$deg.  Without the angular-size restriction completeness falls to 94\%~\citep{3CRR_Survey}.  These estimates are based on the completeness of the 4CT survey, its coverage of the 3C survey area and the number of galaxies in the 4CT survey that are absent from the 3CR survey.  We have used the spectra of the galaxies in the 3CRR survey, without the angular size restriction, to estimate a typical luminosity for RGs.  A completeness of 94\% suggests that there are approximately 10 galaxies absent in the survey and that our average luminosity is well informed.

Of the 173 galaxies in the survey, two had no linear size measurement available and so it was not possible to estimate their spectra.  The average flux density at $178$\,MHz is $28.31\pm6.72$\,Jy without these galaxies and $28.15\pm6.65$\,Jy when including them.  Therefore their exclusion will only affect the average luminosity if their linear sizes are significantly larger than the other galaxies in this catalogue.  Correspondingly, since the error on the average luminosity will scale as~$\sqrt{N}$, where $N$ is the number of galaxies in the sample, the exclusion of these two objects results in a difference of $<1\%$ to the uncertainty in the average luminosity.  

\subsubsection{Geometry considerations}

In \secref{sec:RGs} we used an opening angle of $22.6\,$deg for the geometry of a typical RG based on the average value found in a sample of 362~sources by \citet{Pushkarev2017}.  In order to understand how this parameter affects our estimate of the CRB, we repeat our calculation for opening angles of 10, 25 and 40\,deg.  Varying this parameter has a direct effect on the average path length~$l$ through an RG, and hence on its optical depth~$\tau$: increasing the opening angle will increase $\tau$, and hence increase the radio emission per \eqnref{RG_Intensity}.  It also has an indirect effect, influencing the calculation of the equipartition magnetic field~$B_{\rm eq}$: increasing the opening angle will decrease $B_{\rm eq}$, which will generally decrease the radio emission.

In the optically-thick regime the intensity is independent of $l$ and only weakly dependent on $B_{\rm eq}$ and so the low-frequency emission does not vary greatly with a change in opening angle.  However in the optically-thin regime the intensity from a single galaxy is proportional to $l B^{3.75}$.  Due to the stronger dependence on $B$, the indirect effect is dominant, and increasing the opening angle will cause a net decrease in the intensity of the emission at high frequencies, for a single galaxy.

\begin{figure}
    \centering
    \includegraphics[width=\linewidth]{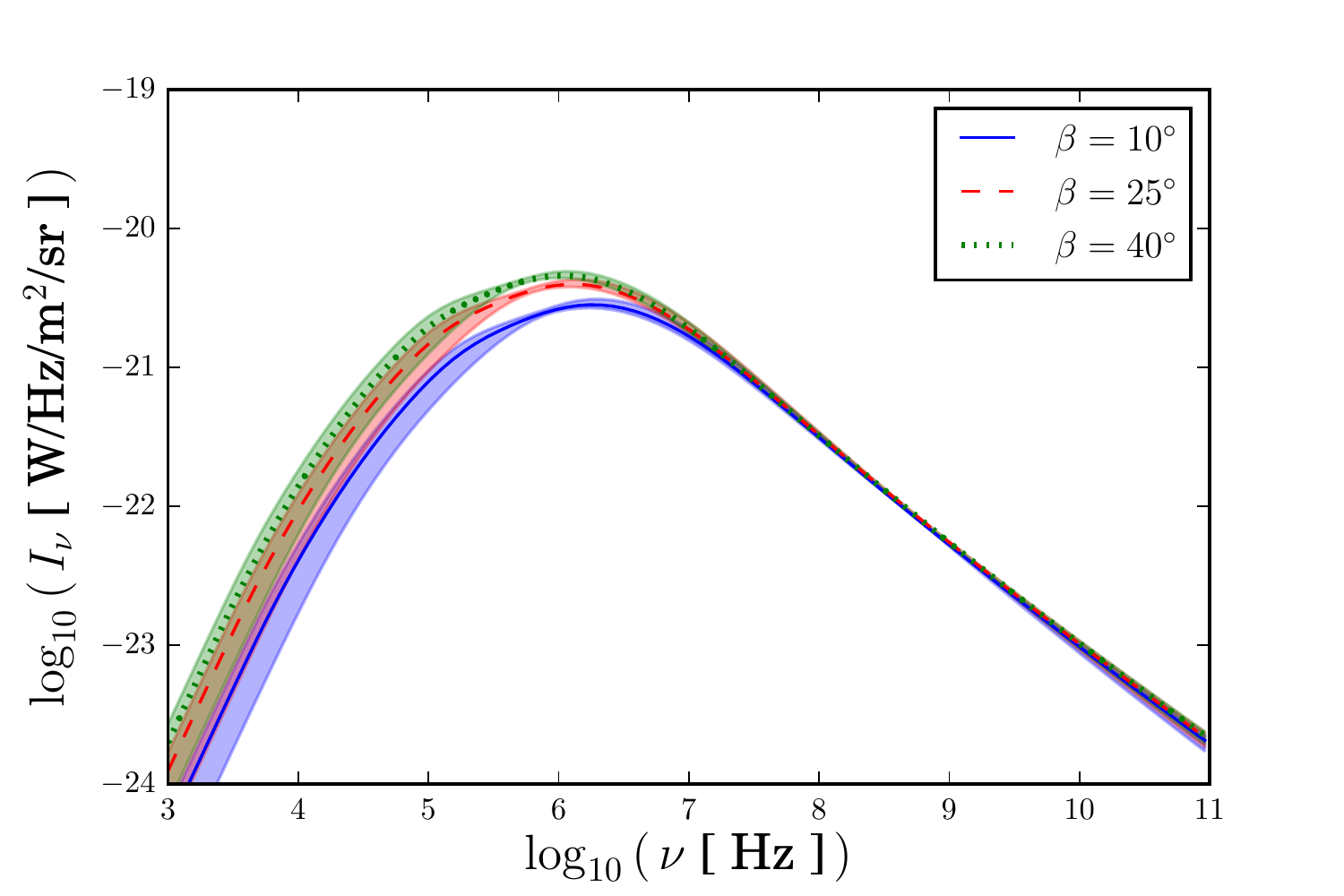}
    \caption{The variation in the RG contribution to the CRB caused by varying the opening angle~$\beta$.}
    \label{fig:alphaVar}
\end{figure}

The variation in the intensity of each galaxy as a function of opening angle propagates through to the average spectrum for the 3CRR survey and consequently through to the CRB contribution (see \figref{fig:alphaVar}).  In \eqnref{Iztotal} the average spectrum, $L_{\nu'}$ is normalized at 1.4\,GHz, in the optically-thin, high-frequency regime.  Consequently, an increase in the assumed opening angle causing a decrease in the high-frequency emission~$L_{\nu'}$ for a single galaxy, as above, will ultimately cause an increase in our calculated CRB at low frequencies.

The maximum difference in the RG contribution to the CRB due to a changed average opening angle between $\beta=40^{\circ}$ and $\beta=25^{\circ}$ is a factor of $\sim 6$, and between $\beta=25^{\circ}$ and $\beta=10^{\circ}$ is a factor of $\sim 9.5$.  These differences are most pronounced at low frequencies ($\nu\sim10$\,kHz) and decrease towards higher frequencies.  As the low-frequency emission is dominated by SFGs (see \figref{fig:TotalBackground}), changing the RG contribution even by such a large factor will not have an overwhelming effect on the total CRB.

\subsubsection{The synchrotron self-absorbed spectral index} \label{sec:disc:SSA}

\citet{Protheroe1996} calculate the spectrum of their typical RG at low frequencies, in the regime in which it is optically thick and subject to synchrotron self-absorption, using a power-law approximation $L \propto \nu^\alpha$ with $\alpha = 2.5$~\citep{Longair}.  In this work, we calculate the spectrum using the source function, $I_{\nu} = \varepsilon^S_{\nu} / \kappa^S_{\nu}$, where $\varepsilon^S_{\nu}$ and $\kappa^S_{\nu}$ are calculated using \eqnrefii{eps_s}{kap_s} respectively.  The spectral indices we recover at the lowest frequencies are typically $\alpha \sim 2$.

Both spectral indices, $\alpha = 2.5$ and $\alpha = 2$, are theoretically justified under different circumstances.  At frequencies below the characteristic emission frequency for electrons at their minimum energy~$E_0$ \cite{Longair},
 \begin{equation}
  \nu_{\rm m} = \bigg( \frac{E_0}{m_{\rm e} c^2} \bigg)^{\!\!2} \bigg( \frac{e B}{2 \pi m_{\rm e}} \bigg),
  \label{eqn:num}
 \end{equation}
the spectral index is $\alpha = 2$, as shown in \appref{app:SSA}.  If this frequency lies below the frequency $\nu_{\rm SSA}$ of the transition between the optically-thick, synchrotron self-absorbed regime and the optically-thin regime, then at frequencies $\nu_{\rm m} < \nu < \nu_{\rm SSA}$ a spectral index of $\alpha = 2.5$ applies.  For further discussion of this point, see \citet{SSA2} (cf.\ their Fig.~1).

\begin{figure}
    \centering
    \includegraphics[width=\linewidth]{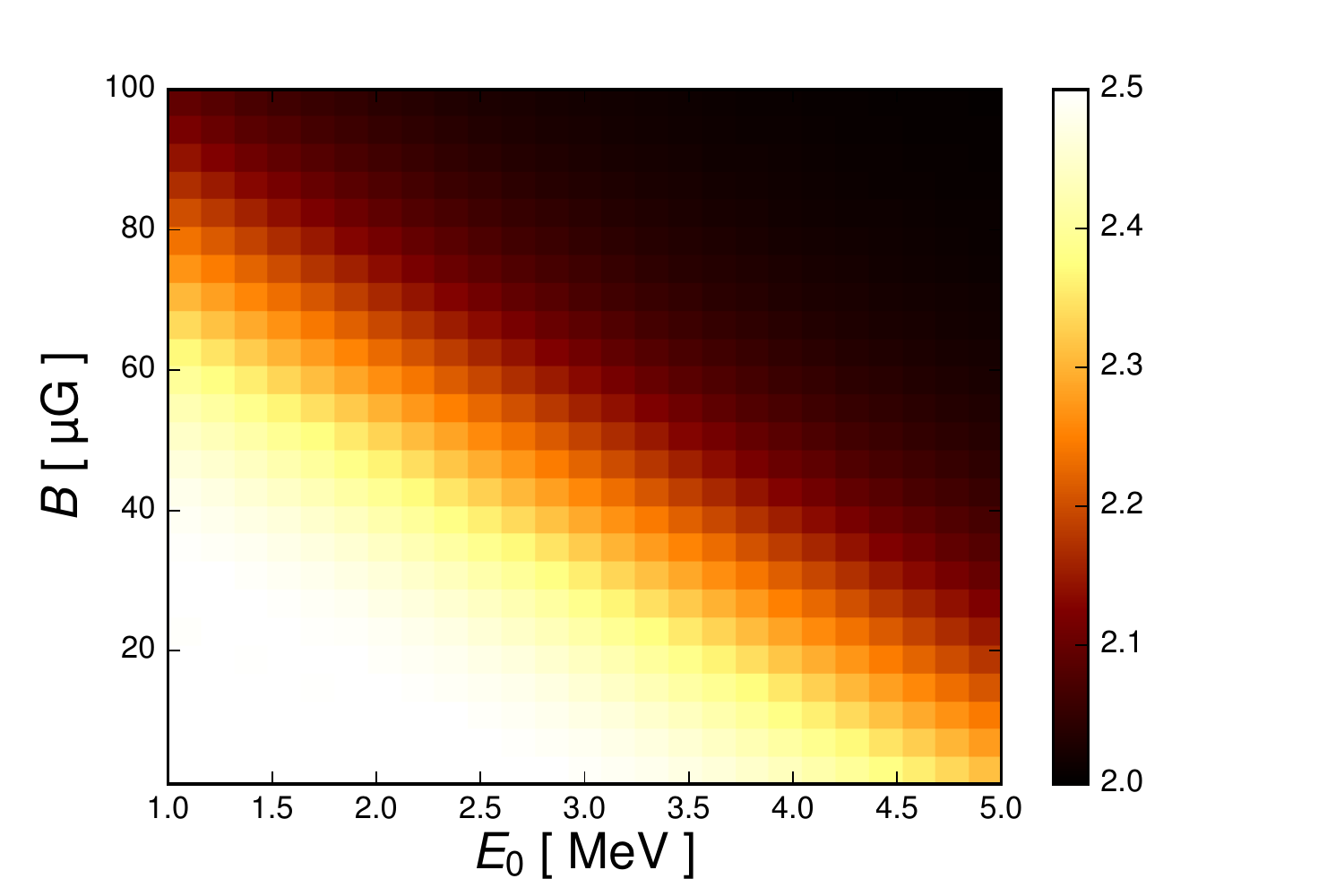}
    \caption{The spectral index between 0.3 and 3\,kHz for the galaxy 4C12.03 as a function of magnetic field strength~$B$ and minimum electron energy~$E_0$.  Note that while the magnetic field strengths are typical of a RG, the cut-off energies shown here are much smaller than the $E_0 = 100$~MeV we used in calculating the contribution of RGs to the CRB. This illustrates that the transition between $\alpha=2$ and $\alpha=2.5$ happens at much lower frequencies than the real characteristic frequency for the galaxy 4C12.03, $\nu_m \sim 0.7$~MHz.}
    \label{fig:BvsE0}
\end{figure}

To illustrate this point, we show in \figref{fig:BvsE0} the low-frequency spectral index for the galaxy 4C12.03 under our model.  There is a transition from $\alpha = 2$ to $\alpha = 2.5$ dependent on $B$ and $E_0$, as expected from \eqnref{eqn:num}.  The range of $B$ shown on this plot is typical for the magnetic field strengths in an RG, but our assumed value of $E_0=100$\,MeV lies well off the plot, deep in the regime in which $\alpha = 2$.  We therefore expect consistent spectral indices of $\alpha = 2$ at low frequencies for this and similar sources, unless the electron population extends to minimum energies $E_0$ much less than our assumed value.  We calculate for 4C12.03 , using our assumed value of $E_0 = 100$~MeV, an equipartition magnetic field of $B_{\rm eq} = 6.39\,\upmu$G and thus $\nu_m \sim 0.7$\,MHz; a spectral index of $\alpha = 2.5$ will apply above this frequency.

\subsection{The low-energy cut-off} \label{Sec:disc:cut}
For both SFGs and RGs, we assumed a hard low-energy cut-off of the electron spectrum.  This is consistent with the~\citet{Protheroe1996} treatment.  The hard low-energy cut-off mostly affects the low-frequency CRB estimates, having little effect on the peaks of the distributions in~\figrefii{fig:SFG_Background}{fig:RG_Background}.  Therefore, the interesting regions for the purpose of this analysis, as discussed below, are not significantly affected by this property.  Moreover, the large uncertainties in the CRB estimate at low frequencies are likely to dominate over the effect of the low-energy cut-off.

We note that there are observations of RGs in which the low-frequency turnover of a spectrum can be best described by a low-energy cut-off in the electron energy density, such as in Cygnus A~\citep{Carilli1991}.  However, this is the subject of some contention and it is argued that a combination of a low-energy cut off, absorption processes and non-uniform magnetic field strength can contribute to accurately modelling the low frequency spectrum~\citep{McKean2016}.

In calculating our average RG spectra, we consider varying levels of magnetic field strength between the 3CRR galaxies.  This gives our average galaxy a non-uniform spectral index and consequently a non-uniform magnetic field.  Additionally, we note that compression and expansion of the plasma in a RG can cause electrons to gain and lose energy as they are injected into the plasma.  This can adjust the low-energy cut-off both across the galaxy and across a sample of galaxies.  However, in a large enough ensemble of galaxies, the electron energy density, on average, can be reasonably approximated by the assumptions made in~\citet{Protheroe1996}.

\subsection{High-redshift contributions} \label{Sec:redshiftlimit}

The most distant galaxy observed is GN-z11 with a spectroscopic redshift of $z=11.09^{+0.08}_{-0.12}$~\citep{Oesch2016}.  This is consistent with our understanding that galaxies began to form during a period of cosmic history known as the epoch of re-ionization (EoR).  Consequently, this epoch can be used to determine an upper limit on redshift for our background calculation,~$z_{\rm EoR}$.  Since the EoR was not an instantaneous process but rather a continuous process over a period of time, this redshift traditionally denotes an approximate mid-point for the EoR.

Determination of $z_{\rm EoR}$ is an active research area with a primary focus towards the global 21\,cm line.  Recent results from the Experiment to Detect the Global EoR Signature (EDGES) places the EoR at $z = 15$--20~\citep{EDGES2018}.  However, we note that the EDGES result has yet to be validated by other experiments and does not agree with previous estimates of $z_{\rm EoR}$.

Re-ionization suppresses small-scale CMB anisotropies as a result of an increased electron density and induces anisotropies in the polarisation of the CMB.  By comparing observational data with equivalent simulations of anisotropies in the event that re-ionization had not taken place it is therefore possible to calculate $z_{\rm EoR}$.  The \citet{Planck2018} used this technique to calculate $z_{\rm EoR} = 7.68 \pm 0.79$ and \citet{Spergel2007} used WMAP to estimate that the EoR started at $z = 11$ and ended at $z = 7$.  Measurements of the Gunn-Peterson trough in the spectra of four quasars from the Sloan Digital Sky Survey put the end of the EoR at $z \sim 6$~\citep{Becker2001}.

\begin{figure}
    \centering
    \includegraphics[width=\linewidth]{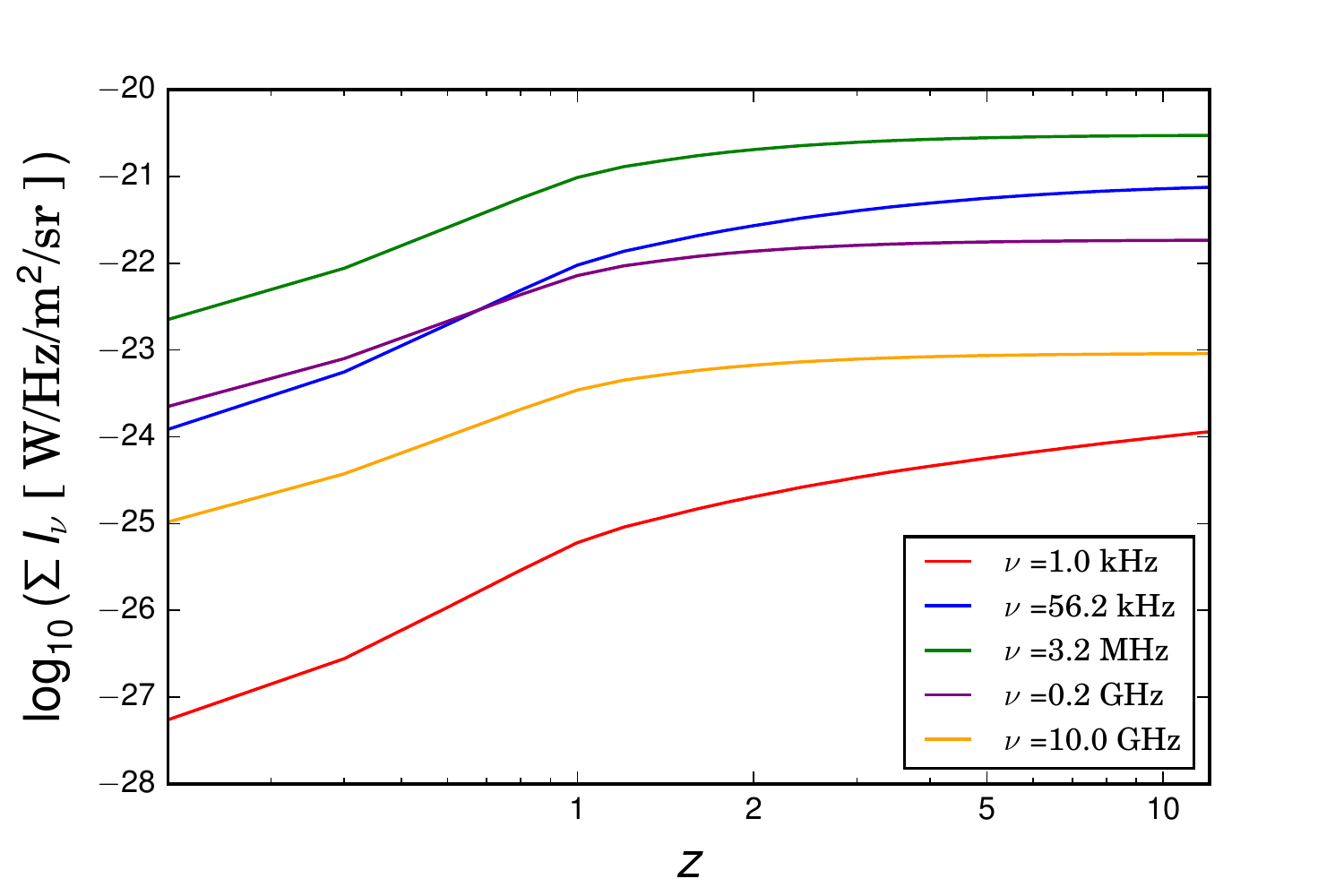}
    \caption{Cumulative contribution to the CRB of RGs at different redshifts~$z$.  At low frequencies there is a substantial contribution from the high-redshift Universe, with concomitant uncertainties associated with early source evolution, but at frequencies $\gtrsim 1$\,MHz the CRB originates primarily within the EoR horizon.}
    \label{fig:RGBackgroundvsZ}
\end{figure}

These measured values of $z_{\rm EoR}$ inform the redshift limit used in \secrefii{sec:SFGs}{sec:RGs}.  To determine the effect of uncertainty in this parameter, we recalculate the integral in \eqnref{Iztotal} for different limits in $z$ to determine the contributions to the CRB arising from different ranges of redshift.  The results of this test for the RG contribution are shown in \figref{fig:RGBackgroundvsZ}; the results for the SFG contribution are similar.  The high-frequency ($\gtrsim 1$\,MHz) component of the CRB originates from redshifts within the EoR horizon, and so is not sensitive to its exact value, but at lower frequencies the CRB is produced primarily by more distant sources, and so the EoR constitutes a significant source of uncertainty.

\section{UHE photons}
\label{sec:intlen}

\subsection{Calculation of UHE photon attenuation length}

UHE photons with energy~$E$ will be attenuated by the field of background photons of energy $\varepsilon = h \nu$, such as the CRB, via electron pair-production with attenuation length~\citep{Protheroe1996}
 \begin{equation}
  \lambda(E) = 8 E^2 \bigg(
    \int_{\varepsilon_{\rm min}}^\infty \! d\varepsilon \, \frac{n(\varepsilon)}{\varepsilon^2}
    \int_{S_{\rm min}}^{S_{\rm max}(\varepsilon,E)} \! ds \, s \, \sigma(s)
  \bigg)^{\!\!\! -1}
 \end{equation}
where
 \begin{equation}
  n(\varepsilon) = \frac{ 4 \pi I_{\nu} }{ h c \varepsilon }
 \end{equation}
is the number density of CRB photons and
 \begin{equation}
  s = 2 \varepsilon E(1 - \cos\theta)
  \label{CoMenergy}
 \end{equation}
is the energy in the centre-of-momentum frame of the two photons interacting at an angle~$\theta$, being integrated between a minimum value of
 \begin{equation}
  S_{\rm min} = (2 m_{\rm e} c^2)^2
  \label{eqn:Smin}
 \end{equation}
defined by the rest-mass energy of an electron-positron pair and a maximum value of
 \begin{equation}
  S_{\rm max}(\varepsilon,E) = 4 \varepsilon E .
 \end{equation}
The minimum background photon energy, from \eqnrefii{CoMenergy}{eqn:Smin}, is
 \begin{equation}
  \varepsilon_{\rm min} = \frac{S_{\rm min}}{4E}
 \end{equation}
and the cross section for photon-photon interactions is~\citep{Jauch}
 \begin{equation}
  \sigma(s) = \frac{3}{16} \sigma_T (1 - v^2)
   \bigg[ (3 - v^4) \ln\bigg( \frac{1+v}{1-v} \bigg) - 2v (2 - v^2) \bigg]
 \end{equation}
where
 \begin{equation}
  v = \bigg(   {1 -} \frac{4(m_{\rm e} c^2)^2 }{ s } \bigg)^{1/2}
 \end{equation}
and $\sigma_T = 6.65 \times 10^{-29}$\,m$^{-2}$ is the Thomson cross section.

UHE photons will also be attenuated by the same background photon field through muon pair-production, which can be calculated similarly with the electron mass replaced with the muon mass.  However, the cross-section for this process is much lower, and it does not contribute significantly to the attenuation.

\subsection{Discussion} \label{sec:disc:attenlen}

\begin{figure}
    \centering
    \includegraphics[width=\linewidth]{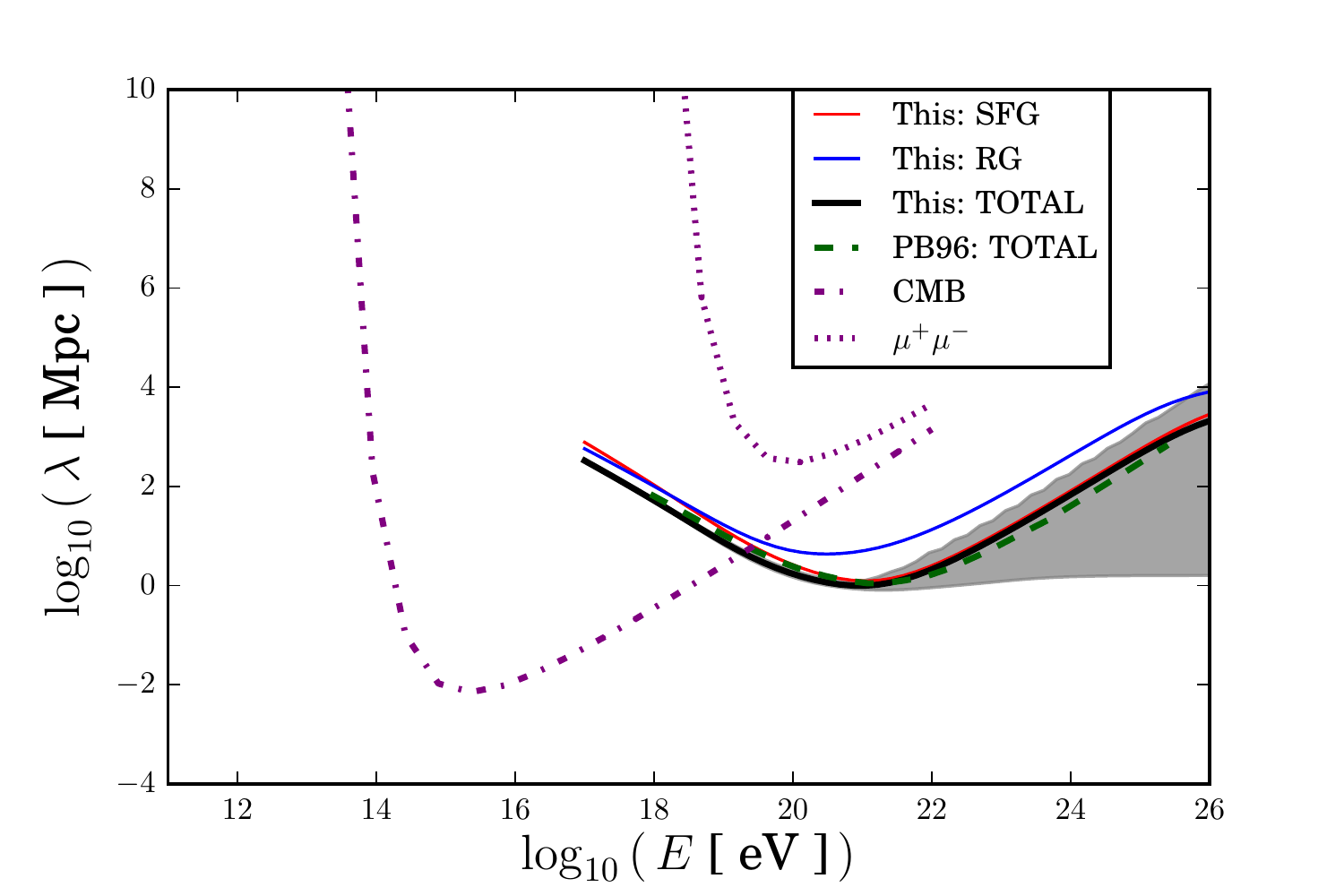}
    \caption{The attenuation length of UHE photons is shown as calculated here and as by \citet{Protheroe1996}.  Also shown is the attenuation length for pair production of $e^-\,e^+$ and $\mu^-\,\mu^+$ pairs resulting from interactions with the CMB.  }
    \label{fig:AttenLen}
\end{figure}

The attenuation length for UHE photons calculated as above, due to the CRB as estimated in this work, is shown in \figref{fig:AttenLen} alongside the equivalent result from \citet{Protheroe1996}.  We also show the attenuation lengths calculated for our CRB estimate resulting from either RGs or SFGs alone.  At energies $\gtrsim 10^{19}$\,eV, where the CMB no longer dominates the attenuation of UHE photons, the attenuation due to the CRB is primarily due to SFGs, with RGs making only a minor contribution.  The uncertainties in the CRB at low frequencies (see \secref{sec:disc}) result in uncertainty in the photon attenuation length at higher energies $\gtrsim 10^{22}$\,eV.  Predictions exist for fluxes of UHE photons resulting from exotic physical processes that extend to these energies~\citep{ellis2006,aloisio2015}, but the cosmogenic photon flux is not expected to do so~\citep{taylor2009,hooper2011}.

\begin{figure}
    \centering
    \includegraphics[width=\linewidth]{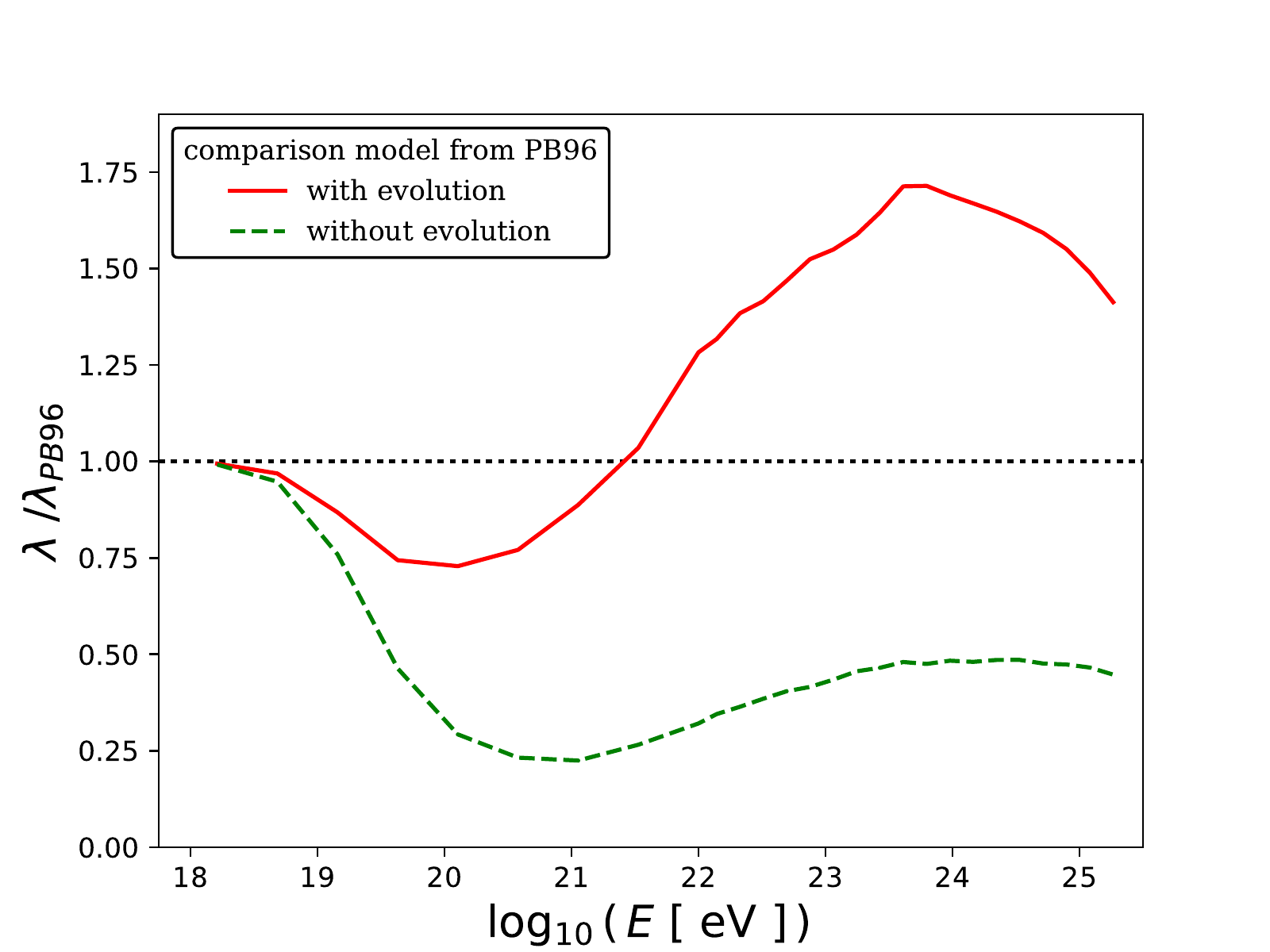}
    \caption{The ratio of predicted UHE photon attenuation length from this work and that of \citet{Protheroe1996}, for both their models with and without the effects of evolution.  All models converge to the same attenuation length at lower photon energies, where the CMB is dominant.}
    \label{fig:AttenLen_Ratio}
\end{figure}

\figref{fig:AttenLen_Ratio} shows the photon attenuation length calculated in this work, including the effects of the CMB, relative to two models from \citet{Protheroe1996}, with and without allowing for cosmological evolution of radio sources (see \secref{sec:comparison}).  Our results are broadly consistent with their model that, like ours, includes evolution, though we do predict stronger attenuation of cosmogenic UHE photons at the expected energies $\lesssim 10^{20}$\,eV.  We predict much stronger attenuation than their model that neglects evolution, at all photon energies for which the CRB is important.

 {These differences in the attenuation length affect the expected flux of UHE photons.  The number of detected UHE photons might be expected to scale as the attenuation length $\lambda$.  While this will not hold strictly due to the uneven distribution of sources, cascading of photons down to lower energies, etc., as a na\"ive estimate it implies that the flux of UHE photons at $10^{20}$\,eV will be a factor $\sim 1.4$ lower than would be predicted from \citeauthor{Protheroe1996}'s with-evolution estimate, or a factor $\sim 3$ lower than predicted from their without-evolution estimate.}

A more rigorous exploration of the implications of our results for the expected number of UHE photons will require implementation of our CRB in Monte Carlo simulations of UHE particle propagation~\citep{armengaud2007,batista2016,aloisio2017}.  To aid such an implementation, we provide a polynomial fit to our CRB estimate,
 \begin{equation}
  \log_{10}(I_\nu / \textnormal{W\,Hz$^{-1}$\,m$^{-2}$\,sr$^{-1}$}) = \sum_i p_i \log_{10}\textnormal{($\nu$/MHz)}^i
  \label{eqn:coeffs}
 \end{equation}
 where the coefficients~$p_i$ parameterising the fit are listed in \tabref{tab:coeffs}.  This parameterisation is accurate to within a 5\% tolerance throughout the displayed frequency range.
 
\begin{table}
 \centering
 \begin{threeparttable}
  \caption{Coefficients for parameterisation of our CRB estimate (\eqnref{eqn:coeffs}).}
  \begin{tabular}{lc}
   \toprule
   Coefficient & Value \\
   \midrule
   $p_{0}$ & $-1.9847 \times 10^{+1}$ \\
   $p_{1}$ & $-2.9857 \times 10^{-1}$ \\
   $p_{2}$ & $-2.6984 \times 10^{-1}$ \\
   $p_{3}$ & $+9.5393 \times 10^{-2}$ \\
   $p_{4}$ & $-4.9059 \times 10^{-2}$ \\
   $p_{5}$ & $+4.4297 \times 10^{-3}$ \\
   $p_{6}$ & $+7.6038 \times 10^{-3}$ \\
   $p_{7}$ & $-1.9690 \times 10^{-3}$ \\
   $p_{8}$ & $-2.2573 \times 10^{-4}$ \\
   $p_{9}$ & $+1.1762 \times 10^{-4}$ \\
  $p_{10}$ & $-9.9443 \times 10^{-6}$ \\
   \bottomrule
  \end{tabular}
  \label{tab:coeffs}
 \end{threeparttable}
\end{table}

\section{Conclusions}
\label{sec:conc}

We have presented an updated estimate of the cosmic radio background and the resulting attenuation length of ultra-high energy photons, building on the work of \citet{Protheroe1996}.  This new estimate also provides associated uncertainties as a function of frequency derived from observational constraints on key physical parameters.  In doing this we have also investigated the variation in the spectrum of the cosmic radio background as a function of these parameters, as well as accounting for the expected variation in spectral index among the population of radio galaxies.  The new estimate presented in this work shows better agreement with observational constraints from source counts than previous calculations.

We estimate a cosmic radio background that is generally more intense than calculated by \citet{Protheroe1996} at the frequencies of interest, and best supports their model including the effects of cosmological evolution.  The resulting attenuation lengths for ultra-high-energy photons, at the energies expected from cosmogenic processes ($< 10^{20}$\,eV), are shorter than previously predicted by a factor up to 3, depending on the energy and specific model for comparison.  This suggests a decrease in the expected number of photons arriving with such energies.

\section*{Acknowledgements}

The authors thank Paddy Leahy for his useful comments on this work.  JDB \& AMS gratefully acknowledge support from the UK Research \& Innovation Science \& Technology Facilities Council (UKRI-STFC) through grant ST/P005764/1.  This research was supported by JBCA, University of Manchester.

\appendix

\section{Radiation theory and emission mechanisms} \label{app:radiation}

Generally, the luminosity can be related to the intensity~$S_{\nu}$ by~\citep{Rohlfs}
 \begin{equation}
  L_{\nu} = 4 \pi d^2 \int_\Omega S_{\nu} \, d\Omega
  \label{L_nu}
 \end{equation}
where $d\Omega$ is the element of solid angle subtended by the galaxy and $d$ is the luminosity distance.

Consider the emissivities (or emission coefficients)~$\varepsilon_{\nu}^{\rm S,F}$ in W\,m$^{-3}$\,Hz$^{-1}$\,sr$^{-1}$ and the absorption coefficients~$\kappa_{\nu}^{\rm S,F}$ in m$^{-1}$ to characterise the synchrotron (S) and free-free (F) radiation processes.  The intensity can be expressed as~\citep{Longair}
 \begin{equation}
  S_{\nu}^{\rm th} =
    \frac{ \varepsilon_{\nu}^{\rm S} }{ \kappa_{\nu}^{\rm F} + \kappa_{\nu}^{\rm S} }
     \, \bigg[ 1 - e^{ -( \tau_{\nu}^{\rm S} + \tau_{\nu}^{\rm F} ) } \bigg]
   +\frac{ \varepsilon_{\nu}^{\rm F} }{ \kappa_{\nu}^{\rm F} }
     \, \bigg[1 - e^{ -\tau_{\nu}^{\rm F} } \bigg]
  \label{Sth}
 \end{equation}
for an SFG, in which synchrotron emission/absorption and free-free absorption are all important, and as
 \begin{equation}
  S_{\nu}^{\rm th} =
    \frac{ \varepsilon_{\nu}^{\rm S} }{ \kappa_{\nu}^{\rm S} }
     \, \bigg[ 1 - e^{ -\tau_{\nu}^{\rm S} } \bigg]
  \label{RG_Intensity}
 \end{equation}
for an RG, in which only synchrotron emission and absorption need to be considered.  

The expression
 \begin{equation}
  \tau_{\nu}^{\rm S,F} = \int_{l} \kappa_{\nu}^{\rm S,F} dl
 \end{equation}
gives the optical depths along the line of sight~$l$ through the medium, quantifying the ratio of incident to transmitted radiant power through the galaxy.  The superscript th, for \emph{theoretical}, is used to differentiate these quantities from the \emph{observational} values used elsewhere in this work.  In the following sections we describe the mathematical expressions for the quantities~$\varepsilon_{\nu}$, $\kappa_{\nu}$ and~$\tau_{\nu}$.

\subsection{Free-free radiation}

The free-free absorption coefficient is given by~\citep{Cane1979}
 \begin{equation}
  \kappa_{\nu}^{\rm F} = 1.64 \times 10^5 \,
  \bigg( \frac{T_{\rm e}}{1\,{\rm K}} \bigg)^{\!\!\!-1.35} \,
  \bigg( \frac{\nu}{1\,{\rm MHz}} \bigg)^{\!\!\!-2.1} \,
  \bigg( \frac{\neth}{1\,{\rm cm}^{-3}} \bigg)^{\!\!2} \,
  \textnormal{pc}^{-1}
  \label{kap_f}
 \end{equation}
where $\neth$ is the number density of electrons of temperature~$T_{\rm e}$.  Similarly, the free-free emissivity can be expressed as~\citep{Rohlfs}
 \begin{equation}
  \varepsilon_{\nu}^{\rm F} = 4.15 \times 10^{-40} \,
  \bigg( \frac{T_{\rm e}}{1\,{\rm K}} \bigg)^{\!\!\!-0.5} \,
  \bigg( \frac{\neth}{1\,{\rm cm}^{-3}} \bigg)^{\!\!2} \, \langle g \rangle \,
  \textnormal{W\,m$^{-3}$\,Hz$^{-1}$\,sr$^{-1}$}
  \label{eps_f}
 \end{equation}
where the Gaunt factor~\mbox{$\langle g \rangle$} is a relativistic correction and is given by
 \begin{equation}
  \langle g \rangle = 
  \begin{cases}
   \ln\! \bigg[ 0.05 \, \bigg( \frac{\nu}{1\,{\rm GHz}} \bigg)^{\!\!\!-1} \bigg]
    + 1.5 \, \ln\! \bigg( \frac{T_{\rm e}}{1\,\textnormal{K}} \bigg) \\
   1, \:\: \textnormal{for} \: \frac{\nu}{1\,{\rm MHz}} \gg \bigg( \frac{T_{\rm e}}{1\,{\rm K}} \bigg)^{\!\!1.5} .  \\
  \end{cases}
 \label{g}
 \end{equation}

It is worth noting that free-free emission as described by \eqnrefii{eps_f}{g} has a low-frequncy cut-off, being suppressed below the plasma frequency~\citep{Longair}
 \begin{equation}
  \nu_{\rm plasma} = \frac{1}{2\pi}\,\sqrt{\frac{\neth\,e^2}{\varepsilon_0\,m_{\rm e}}},
  \label{nuplasma}
 \end{equation}
where $e$ is the electron charge and $\varepsilon_0$ is the vacuum permittivity.  

\subsection{Synchrotron radiation}

The emissivity of synchrotron radiation can be expressed as~\citep{Pacho}\footnote{Note that Eq.~5 of \citep{Protheroe1996} is from Eq.~3.39 of \citep{Pacho}, for synchrotron emission in a single polarisation.  We use Eq.~3.40 of \citep{Pacho}, for the total synchrotron emission.}
 \begin{equation}
  \varepsilon_{\nu}^{\rm S} = \, c_3 \, B \, \langle\sin\theta\rangle
   \int n_{\rm e}(E) \, F(x) \, dE
  \label{eps_s}
 \end{equation}
where $n_{\rm e}(E)$ is the electron energy density (or spectrum of cosmic-ray electrons), $B$ is the magnetic field strength, $\langle\sin{\theta}\rangle = 0.785$ for isotropic electrons and $c_3 = 1.87 \times 10^{-23}$\,esu$^3$\,g$^{-1}$cm$^{-2}$\,s$^2$.  The function in the integral is defined as
 \begin{equation}
  F(x) = x \int_{x}^{\infty} \! K_{5/3}(z) \, dz
  \label{Fx}
 \end{equation}
where $K_{5/3}$ is a modified Bessel function of the second kind, and
 \begin{equation}
  x = \bigg( \frac{2}{3} \langle\sin\theta\rangle^{-1} \gamma^{-2} \bigg)
   \, \frac{\nu}{\nu_g},
 \end{equation}
where $\gamma = E/(m_{\rm e}\,c^2)$ is the Lorentz factor for electrons of energy~$E$ and mass~$m_{\rm e}$, and
 \begin{equation}
  \nu_g = \frac{e \, B}{2 \, \pi \, m_{\rm e}}
 \end{equation}
is the cyclotron frequency for an electron.

Similarly, the synchrotron absorption coefficient is given by~\citep{Protheroe1996}
 \begin{equation}
  \kappa_{\nu}^{\rm S} = \frac{c^2 \, c_3}{2 \, \nu^2} \, B \langle\sin\theta\rangle
   \int\! E^2 \, \frac{d}{dE} \bigg[ \frac{n_{\rm e}(E)}{E^2} \bigg] \, F(x) \, dE,
  \label{kap_s}
 \end{equation}
where $c$ is the speed of light and the other symbols have been defined previously.

\section{Synchrotron self-absorbed spectral index} \label{app:SSA}

For synchrotron emission in the optically-thick, self-absorbed regime the intensity of the spectrum is determined by the source function.  From \eqnrefii{eps_s}{kap_s} we see that this is given by
 \begin{align}
  I_{\nu}
   = \frac{ \varepsilon_{\nu}^{\rm S} }{ \kappa_{\nu}^{\rm S} }
    &= -\frac{2\,\nu^2}{c^2}
     \frac{
      \int\! dE \, n_{\rm e} \, F(x)
     }{
      \int\! dE \, E^2 \, \frac{d}{dE} [n_{\rm e}/E^2] \, F(x)
     } \\
    &= -\frac{2\,\nu^2}{c^2}
     \frac{
      \int\! dE \, n_{\rm e} \, F(x)
     }{
      \int\! dE \, F(x) \, (n_{\rm e}' - 2 n_{\rm e} / E)
     },
 \end{align}
where $n_{\rm e}'$ denotes the derivative of the electron energy density with energy.  The intensity then becomes
 \begin{align}
  I_{\nu} = -\frac{2\,\nu^2}{c^2}
   \frac{
    \int\! dE \, n_{\rm e} \, F(x)
   }{
    \int\! dE \, F(x) \, n_{\rm e}' - 2 \int\! dE F(x) ( n_{\rm e} / E )
   }
 \end{align}
and, using integration by parts over the interval in $E$ from 0 to $\infty$,
 \begin{align}
  I_{\nu} = -\frac{2\,\nu^2}{c^2}
   \frac{
    \int\! dE \, n_{\rm e} \, F(x)
   }{
    [F(x)\,n_{\rm e}]^{\infty}_{E=0}
    - \int\! dE \, \frac{dF(x)}{dE} \, n_{\rm e}
    - 2 \int\! dE \, F(x) \, (n_{\rm e}/E)
   } .
 \end{align}
If we assume that the electron population has a minimum energy cut-off at energy $E_0>0$, then $n_e(E=0)=0$; and as $E\rightarrow\infty$, $\gamma\rightarrow\infty$ and hence $F(x)\rightarrow0$.  The first term in the denominator is therefore zero at both limits, and can be neglected.  Noting that $\gamma = E / mc^2$, and all other variables in the expression for $F(x)$ are independent of $E$, this allows us to calculate,
 \begin{align}
  \frac{\partial F(x)}{\partial E}
   &= \frac{d\gamma}{dE} \frac{\partial F(x)}{\partial \gamma}, \\
   &= \frac{1}{m \, c^2} \frac{-2}{3} \frac{F(x)}{\gamma}, \\
   &= -\frac{2}{3} \frac{F(x)}{E}
  \label{eqn:Fxderiv}
 \end{align}
using the relation $F(x) \propto \gamma^{-2/3}$ from \eqnref{Fx} in the limit $x \ll 1$.

Consequently the second term in the denominator becomes
 \begin{align}
  \int\! dE \, \frac{dF(x)}{dE} \, n_{\rm e}
   = -\,\frac{2}{3} \int\! dE \, F(x) \, (n_{\rm e}/E)
 \end{align}
and the source function reduces to
 \begin{align}
  I_\nu =
   \frac{3 \, \nu^2}{2 \, c^2} \,
   \frac{ \int\! dE \, n_{\rm e} \, F(x) }{ \int\! dE \, F(x) \, (n_{\rm e}/E) } .
 \end{align}
Finally, noting that $F(x) \propto \gamma^{-2/3} \propto E^{-2/3}$ we have
 \begin{align}
  I_{\nu} =
   \frac{3 \, \nu^2}{2 \, c^2} \,
   \frac{ \int\! dE \, n_{\rm e} \, E^{-2/3} }{ \int\! dE \, n_{\rm e} \, E^{-5/3} } .
  \label{SourceFunc}
 \end{align}
Using this result we demonstrate that a synchrotron self-absorbed spectral index of~2 can arise from a power law electron energy density spectrum as detailed in the following subsection.  Note that $n_{\rm e}$ appears in both the numerator and denominator of \eqnref{SourceFunc} and so its normalisation can be neglected.

\subsection{Power-law electron energy distribution}
\label{sec:power-law}

Consider a population of electrons with a power-law energy distribution, as assumed in this analysis of the CRB, with number density
 \begin{equation}
  n_e(E) =
   \begin{cases}
    E^{-p} & \textrm{for $E > E_0$} \\
    0 & \textrm{elsewhere,}
   \end{cases}
 \end{equation}
and taking $p > 2$ to ensure that the total energy in this population is finite.  In this case, \eqnref{SourceFunc} gives us
 \begin{align}
  I_\nu
   &= \frac{3 \, \nu^2}{2 \, c^2} \frac{
    \int_{E_0}^\infty dE \, E^{-p-2/3}
   }{
    \int_{E_0}^\infty dE \, E^{-p-5/3}
   } \\
   &= \frac{3 \, \nu^2}{2 \, c^2}
    \left( \frac{p + 2/3}{p - 1/3} \right) \, E_0
  \label{eqn:power-law}
 \end{align}
for the source function and hence for the emission of an optically-thick source.  Note the $\propto \nu^2$ Rayleigh-Jeans spectrum, as for purely thermal emission.


\bibliographystyle{elsarticle-num-names}
\bibliography{journals,refs}

\end{document}